\documentclass[reprint, amsmath,amssymb,aps, prb, showkeys,showpacs, superscriptaddress, nofootinbib, floatfix] {revtex4-2}
\usepackage[colorlinks=true,breaklinks=true,allcolors=blue]{hyperref}
\usepackage{physics}
\usepackage{comment}
\usepackage{tabularx}

\newcommand{\wt}[1]{\widetilde{#1}}

\newcommand{\wh}[1]{\widehat{#1}}
\newcommand{\uR}{\widehat{\tau}}

\usepackage{tikzit}
\usepackage{array}
\usetikzlibrary{arrows}

\usepackage{mathtools,tikz-category}
\usepackage{tikzscale}
\usepackage{graphicx}
\usepackage{dcolumn}
\usepackage{bm}

\begin{document}

\title{Lattice Topological Defects in Non-Unitary Conformal Field Theories }
\author{Madhav Sinha}
\email{ms3066@physics.rutgers.edu}
\affiliation{Department of Physics and Astronomy, Rutgers University, 126
Frelinghuysen Rd., Piscataway NJ 08854, USA
}

\author{Thiago Silva Tavares}

, 

\affiliation{Departamento de F\'{i}sica, Universidade Federal de S\~{a}o Carlos, S\~{a}o Carlos, SP 13565-905, Brazil}
\author{Hubert Saleur}
\affiliation{Institut de physique théorique, CEA, CNRS, Université Paris-Saclay, France}
\affiliation{Physics Department, University of Southern California, Los Angeles, USA}
\author{Ananda Roy}
\email{ananda.roy@physics.rutgers.edu}
\affiliation{Department of Physics and Astronomy, Rutgers University, 126
Frelinghuysen Rd., Piscataway NJ 08854, USA
}


\begin{abstract}
Topological defects play a fundamental role in the investigation of symmetries in quantum field theories. For conformal field theories in two space-time dimensions, it is possible to construct these defects using lattice models allowing ab-initio analytical and numerical computations of their characteristics. 
In this work, topological defects are investigated in non-unitary conformal field theories using appropriate variations of the restricted solid-on-solid models. The relevant impurity models and the corresponding defect operators are constructed for the lattice system. Numerical computations are performed for the energy spectrum, eigenvalues of the defect operators as well as thermodynamic characteristics and compared with analytical predictions. Finally, renormalization group flows between the different fixed points are analyzed using numerical methods. 
\end{abstract}

\maketitle

\section{Introduction}
\label{sec:intro}
 Defects in conformal field theories~(CFTs) in two space-time dimensions contain signatures of the characteristics of the underlying theory. Topological defects are those that preserve the continuity of the stress-energy tensor across the defect~\cite{Petkova:2000ip, Frohlich2004}. These defects correspond to symmetries which do not necessarily obey the group-like composition law and can even be non-invertible~\cite{Gaiotto:2014kfa, Cordova:2022ruw, Seiberg:2023cdc, Schafer-Nameki:2023jdn}. The investigation of these defects is important for a large number of physical problems ranging from impurity scattering in condensed matter physics~\cite{Affleck1995conformal, Saleur1998} to string theory~\cite{Polchinski1995, Douglas:1999vm}. 
 
  In contrast to their higher dimensional counterparts, some topological defects in 2D CFTs have explicit lattice realizations~\cite{Aasen2016, Aasen:2020jwb, Belletete2020, Sinha:2023hum, Belletete2023, Roy2024, Sinha:2025jhh}. For many CFTs, the corresponding lattice models are even integrable, leading to analytical characterization of their characteristics. For these models, the topological nature of several defects is manifest already in the lattice incarnation, enabling verification of their properties using small-scale numerical computations without sophisticated analysis of the scaling limit. These have led to characterization of not only thermodynamic properties for the different defect CFTs, but also renormalization group~(RG) flows between them~\cite{Kormos:2009sk, Tavares:2024vtu}. At the same time, the manifestation of topological symmetries directly in the lattice model makes feasible investigation of these symmetries on small-size solid-state engineered quantum systems~\cite{Roy:2024xdi}. In fact, signatures of these symmetries for the simplest case of the Ising CFT have already been obtained on a noisy quantum simulator~\cite{Lamb:2025rbl}. 
  
  \begin{figure}
  \centering
    \includegraphics[width=\linewidth]{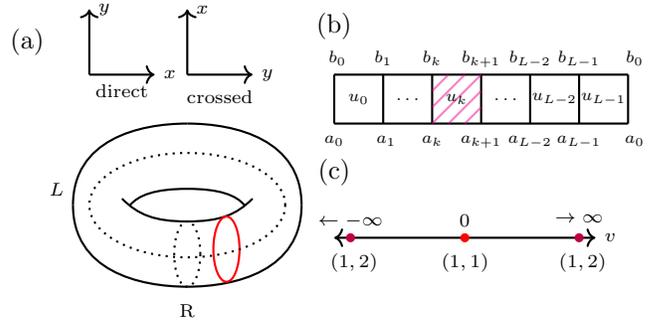}
    \caption{\label{fig:combined-tikz}(a) Schematic of a topological defect (in red)  for a 2D CFT with periodic boundary conditions. In the `direct' channel, this defect is analyzed using a Hamiltonian with a spatially-localized impurity. In the `crossed' channel, the defect manifests itself as an operator that acts on the Hilbert space of the CFT without defects. Here, $x$ is the spatial direction along the chain and $y$ is the imaginary time direction.
    (b) Schematic of the transfer matrix of the associated classical statistical mechanics model with periodic boundary condition. The case of a single defect corresponds to choosing one of the inhomogeneity parameters as~$u_k = u + iv$, keeping all the others~$u$. The Hamiltonian~[Eq.~\eqref{eq:def_Ham_simp}] is obtained in the extreme anisotropic limit. 
    (c) Schematic for the different topological defect fixed points as a function of the defect parameter $v$.}
    \end{figure}
  While topological defects have been extensively analyzed using lattice models of unitary minimal 2D CFTs, the problem has been less explored for their non-unitary counterparts. Unitarity is deeply intertwined with foundational aspects of QFTs and its absence leads to  fundamentally different QFT phenomena absent in unitary models. Notable examples include spectral singularities known as exceptional points~\cite{miri2019exceptional, Bergholtz2021, ding2022non}, a larger class of symmetry protected topological phases~\cite{Altland1997, Ludwig_2016, Gong2018, Kawabata2019, Zhou2019} as well as novel band structure features~\cite{Hatano1996, Shen2018} such as point gap topology~\cite{Kawabata2019, Gong2018, weidemann2020topological, Wang:2020gqg}. For 2D CFTs, non-unitary models have long been of interest. For instance, the lack of applicability of the Zamolodchikov~$c$-theorem~\cite{Zamolodchikov1986} leads to remarkable RG flows~\cite{Fendley1993, Nakayama:2024msv, Klebanov:2022syt, Katsevich:2025ojk}. Furthermore, non-unitary CFTs can have continuous spectra~\cite{Ikhlef2012, Bazhanov:2020uju} relevant for black hole physics~\cite{Dijkgraaf:1991ba,Maldacena:2000kv}.

The purpose of this work is to investigate lattice Hamiltonians that realize topological defects in non-unitary minimal 2D CFTs in the scaling limit. This is accomplished by using the Temperley-Lieb~(TL) Hamiltonians for restricted solid-on-solid~(RSOS) models~\cite{Andrews:1984af, Saleur:1990uz} generalized to the non-unitary domain~\cite{Ardonne_2011}. The derived models remain integrable and the TL Hamiltonians are straightforwardly expressed in the terms of the~${\rm SU(2)}_k$ S-matrix elements~(see below). Analogous to their unitary counterparts~\cite{Tavares:2024vtu, Sinha:2025jhh}, a parameter-dependent Hamiltonian is constructed from the associated transfer matrix~[Fig.~\ref{fig:combined-tikz}(b)]. Tuning the parameter~($v$) interpolates between the identity and the Kramers-Wannier (KW) defects~[Fig.~\ref{fig:combined-tikz}(c)]. The signatures of the defect are captured in both the direct and crossed channels~[Fig.~\ref{fig:combined-tikz}(a)]. In the former case, the energy spectrum of the impurity Hamiltonians and the thermodynamic entropy are computed. For the crossed channel, the expectation values of the associated defect line operators are obtained. The numerical results are compared with analytical predictions where possible. 

The paper is organized as follows. In Section~\ref{sec:model}, the relevant formulas for the quantum RSOS model and the associated transfer matrices are summarized. In Section~\ref{sec:res_fixed_pt}, the low-lying eigenstates of the different impurity Hamiltonians are computed alongside their corresponding momenta eigenvalues. The numerical results are compared with CFT predictions obtained from the corresponding twisted partition functions. Corresponding results for eigenvalues of the lattice defect operators are also presented. In Section \ref{sec:res_flows}, RG flows are analyzed by computing the variation of the thermodynamic entropy as well as the ground-state energy. A concluding summary is provided in Sec.~\ref{sec:concl}. Appendix \ref{sec:any-chain} presents an alternate formulation of non-unitary RSOS models based on anyonic chains defined using Galois-conjugated fusion categories.

\section{Lattice models for non-unitary CFTs with defects}\label{sec:model}

In this section, the basics of the non-unitary variants of the RSOS models~\cite{Andrews:1984af, Ardonne_2011, tartaglia2016fused} are summarized. The relevant quantum  Hamiltonians, which are the first conserved charges constructed from the transfer matrices, are written in terms of the TL operators. These are defined below. 

For the quantum ${\rm RSOS}(m,m')$ chain of length $L$, the Hilbert space comprises states of $\ket{x_0, x_1, \ldots, x_{L - 1}}$, where~$x_i$ is the height at a given site. Here,~$L$ is taken to be even and periodic boundary conditions are imposed.  The heights are constrained by the rule that two nearest neighboring heights are also neighbors on the Dynkin diagram. 
For the sake of brevity, only those belonging to the A$_{m'-1}$ family are considered. For these models, $x_i \in \{ 1, 2, \ldots , m'-1\}$. The relevant quantum Hamiltonians as well as the transfer matrices for the statistical mechanics model are defined in terms of generators of the TL algebra. The latter~$e_i$, $i = 0,\ldots, L - 1$, satisfy the relations: 
\begin{eqnarray}\label{eq:TL_gen}
 e_j^2 &=& (q + q^{-1}) \, e_j \,,  \nonumber \\[0.2cm]
e_ie_{i\pm1}e_i &=& e_i \, , \\[0.2cm]
e_i e_j &=& e_j e_i \hspace{1cm} |i-j|\geq2 \nonumber \, ,
\end{eqnarray}
where $q = {\rm e}^{{\rm i}a \pi/m'}$ and  $a = m' - m$. The matrix elements of these TL operators are given by:
\begin{equation}
\label{eq:TL-op}
      \bra{\{x'_i\}} e_{i}\ket{\{x_{i}\}} = 
      (\prod_{j \neq i } \delta_{x'_j,  x_j}) \delta_{x_{i-1},x_{i+1}} \frac{ \sqrt{S_{a,x_i}} \sqrt{S_{a,x'_i}}}{S_{a,x_{i-1}}}    \, ,
\end{equation}
where $S$ is the corresponding~${\rm SU(2)}_k$ S-matrix~\cite{DiFrancesco:1997nk}:
\begin{equation}\label{eq:S-mat-WZW}
    S_{j_1,j_2} = \sqrt{\frac{2}{m'}}\sin \left( \frac{ j_1 j_2 \pi }{m'}\right) \, . 
\end{equation}

The defect operator, as well as the computation of the momentum eigenvalues for a given state in the energy spectrum, require the shift operator, $\uR$. The latter is defined as
\begin{equation}\label{eq:shift-op}
 \uR\ket{x_0,x_1,x_2, ......,x_{L-1}} = \ket{x_{L-1},x_0,x_1, ......,x_{L-2}} \, .  
\end{equation}
It satisfies
\begin{align}
    \label{eq:shift-ei-rel}
    \uR e_i &= e_{i+1} \uR,\\
    \label{eq:prod_of_ei_rel}
    \uR^2 e_{L-1} &= e_1  e_2.....e_{L-1}.
\end{align}
Together with~$\{e_i\}$, the operators $\uR$, $\uR^{-1}$ generate the affine TL algebra - aTL$_{L}(q)$~\cite{Belletete2018}. The Hamiltonian governing the quantum ${\rm RSOS}(m,m')$ chain with periodic boundary conditions and without any defect is given by:
\begin{equation}\label{eq:Ham-11}
    H_{0} = - \frac{\gamma}{\pi \sin \gamma} \sum_{i = 1}^{L} e_i \, ,
\end{equation}
where $\gamma  = a \pi/m'$  is the parameter that determines which CFT is reached in the scaling limit. Since $-e_i$ also satisfies the Temperley-Lieb relations but with the scale factor $ \tilde{q} + \tilde{q}^{-1}$, where $\tilde{q} = {\rm e}^{{\rm i}m \pi/m'}$~(recall~$L$ is even), the Hamiltonian~$-H_0$ realizes the RSOS$(m'-m,m')$ model~\footnote{Here we are assuming that $m \neq m'- 1$, as RSOS$(m,m')$ always has $m>1$.}. In this work, by suitable choice of the integers $m,m'$, all the non-unitary minimal models are analyzed, instead of explicitly considering~$-H_0$ for a given choice of~$m, m'$, as was done in \cite{Ardonne_2011}. 

 Note that for $m' - m = 1$, the $e_i$ operators are Hermitian, and so is the Hamiltonian in Eq.~\eqref{eq:Ham-11}. In the scaling limit, these models flow to the unitary minimal model~${\cal M}(m',m'-1)$. In contrast, for $m' - m > 1$, the $e_i$ operators are not Hermitian, even though they are  symmetric, i.e. $e_i^{T} = e_i$. As a result, the TL generators and the corresponding Hamiltonians are complex symmetric matrices. For this choice of~$m'-m$, in the scaling limit, the RSOS model gives rise to the non-unitary ${\cal M}(m',m)$ minimal model CFT. Here, recall gcd$(m,m') = 1$ and $2 \leq m  < m'$, see Appendix~\ref{sec:any-chain} for more. 

 The above Hamiltonian~($H_0$) corresponds to the model without any defects or equivalently, the trivial/identity/(1,1) defect. Here and in what follows, the Kac labels for the topological defects are used to denote them. In order to probe the KW defect and the RG flow emanating from the associated fixed point,~$H_0$ is modified between sites~$k, k+1$ to give rise to the defect Hamiltonian~$H_D$, defined as:
 
\begin{align}\label{eq:def_Ham_simp}
      H_D (v) &=  H_0  +\frac{1}{\sin \gamma} f(v) \,  e_k e_{k+1}\nonumber\\&\qquad  + \frac{1}{\sin \gamma} f(-v) \,  e_{k + 1} e_k \, ,
\end{align}
where 
\begin{equation}\label{eq:f-def-ham}
    f(v) = \frac{\sin ({\rm i}v)}{\sin (\gamma + {\rm i}v)}  \, ,
\end{equation}
see Ref.~\cite{Sinha:2025jhh} for a derivation. For~$v = 0$, the Hamiltonian~$H_0$ is recovered. But, in the limit~$v\rightarrow\infty$, the Hamiltonian~$H_D$ realizes the KW~[$(1,2)$] defect for the corresponding CFT. By varying the parameter~$v$, the RG flow between the KW~$(1,2)$ and the identity~$(1,1)$ defect fixed points can analyzed. In Sec.~\ref{sec:res_fixed_pt}, the low-lying energy spectra are computed to provide numerical evidence for the corresponding lattice models. In order to unambiguously identify each state in the corresponding CFT, the eigenvalues of the corresponding momentum operator are also computed. For the lattice model described above, the momentum operator is given by:
\begin{equation}\label{eq:transl_op}
    U_{k}(v) = \uR  \,  \left[ \mathbf{1} + \frac{\sin ({\rm i}v) }{\sin(\gamma - {\rm i}v)}e_{k} \right] \,    . 
\end{equation}
 It can be checked that~$[U_k(v), H_D(v)]=0$. For $v = 0$ this operator is $\uR$ and for $v \to  \infty$, it is, up to an overall phase, the operator $\uR g_k$. Here,~$g_k$ is the braid operator. The latter are defined as:
 \begin{equation}\label{eq:g_def}
g_j = (-q)^{1/2}\mathbf{1} + (-q)^{-1/2}e_j,
\end{equation}
where $\mathbf{1} $ is the identity operator and  $j = 0, \ldots, L-1$. These braid operators satisfy
\begin{align}
 g_i g_{i+1} g_i &= g_{i+1} g_i g_{i+1} \, ,\\
 g_i g_j &= g_j g_i \hspace{1cm}   \lvert \ i - j \ \rvert \geq 2 \, .
\end{align}

 The impurity Hamiltonians~$H_D(v)$ describe the topological defects for specific choices of~$v$ in the so-called `direct' channel. An equally valid description is in the~`crossed' channel, where the defect acts as an operator on the states of the corresponding CFT with periodic boundary conditions, see panel (a) of Fig.~\ref{fig:combined-tikz}. The relevant defect operators can be constructed for the RSOS$(m, m')$ models using the TL operators. To that end, define the row-to-row transfer matrix~$T \left(  \{ u \} \right)$~[Fig.~\ref{fig:combined-tikz}(b)] as:
 \begin{align}\label{eq:tmat_TLgen_KP}
  T \left(  \{ u \} \right)   &=  \frac{\sin u_0 }{\sin^{L} \gamma } \left(\prod_{j = 1}^{L-1}\tilde{R}_j(u_j) \right)  \uR^{-1} \nonumber
  \\
  &\quad +\frac{\sin \left( \gamma - u_0\right)}{\sin^{L} \gamma} \uR \prod_{j = 1}^{L-1}R_{L-j}(u_{L-j})   \, , 
 \end{align} 
where
 \begin{equation}\label{eq:R&Rtop}
 \begin{split}
    R_j(u_j) = \sin(\gamma - u_j) \, \mathbf{1} + \sin(u_j) \, e_j , \\ \tilde{R}_j(u_j) =   \sin(\gamma-u_j) \,  e_j + \sin(u_j) \, \mathbf{1}  \, .
    \end{split}
\end{equation}
Explicitly, the products in Eq.~\eqref{eq:tmat_TLgen_KP} are given as:
\begin{equation}
\begin{split}
   & \prod_{j = 1}^{L-1} R_{L-j}(u_{L-j})  \equiv R_{L-1}(u_{L-1}) \,  \, \dots \, R_{1}(u_1) \, , \\\
 & \prod_{j = 1}^{L-1}\tilde{R}_j(u_j)  \equiv \tilde{R}_{1}(u_1) \, \tilde{R}_{2}(u_2) \, \dots \,  \tilde{R}_{L-1}(u_{L-1}) \, .
\end{split}
\end{equation}
 The Hamiltonian~$H_D(v)$ commutes with the transfer matrix and is related to it through its logarithmic derivative, see for example, Ref.~\cite{Baxter2013} for more details. For~$u_j = 0\ \forall\ j$, the transfer matrix reduces to the shift operator~$\uR$. On the other hand, for~$u_j = {\rm i}v\ \forall\ j$ and in the limit~$v\rightarrow\infty$, the transfer matrix, up to a scale factor, turns into the topological defect/hoop operators~$Y, \overline{Y}$~\cite{Feiguin:2006ydp, Belletete:2018eua,Belletete2023}. The latter are given by:

\begin{align}
\label{eq:Yoperator_1}
        Y &= (-q)^{-1/2} \,   \, g^{-1}_{1} \ldots g_{L - 1}^{-1}  \uR^{-1} \nonumber\\&
        \qquad +  (-q)^{1/2} \, \uR \, g_{L -1}  \ldots g_1    \,  ,\\
        \label{eq:Yoperator_2}
        \overline{Y} &=  (-q)^{-1/2} \,  \uR  \, g^{-1}_{L - 1} \ldots g_{1}^{-1}\nonumber\\&\qquad +  (-q)^{1/2} \, g_1 \ldots g_{L -1}  \, \uR^{-1}    \,  .
\end{align}

The $Y$ and $\overline{Y}$ operators lie in the center of ${\rm aTL}_{L}(q)$. Note that $Y$ and $\overline{Y}$ are different in general, for instance in D-type RSOS model~\cite{Sinha:2023hum}, but they are the same for A-type RSOS model. Note, similar operators have been discussed in Ref.~\cite{pearce2025unitary} in the context of (1,2) Verlinde lines in non-unitary CFTs. The action of this defect operator on the right eigenstates of the Hamiltonian~$H_0$~(recall that~$H_0$ is non-Hermitian, see below for more) is the lattice counterpart of the action of the CFT defect operator on the corresponding states in the different conformal towers. Indeed, the latter are eigenstates of the corresponding defect operator with eigenvalues given by ratios of certain S-matrix elements~\cite{Petkova:2000ip}. They also serve as diagnostics for the different RG fixed points and are verified using numerical computations. These are described next. 

\section{Results: Fixed Points} \label{sec:res_fixed_pt}
\begin{figure}
\includegraphics[width = 0.5\textwidth]{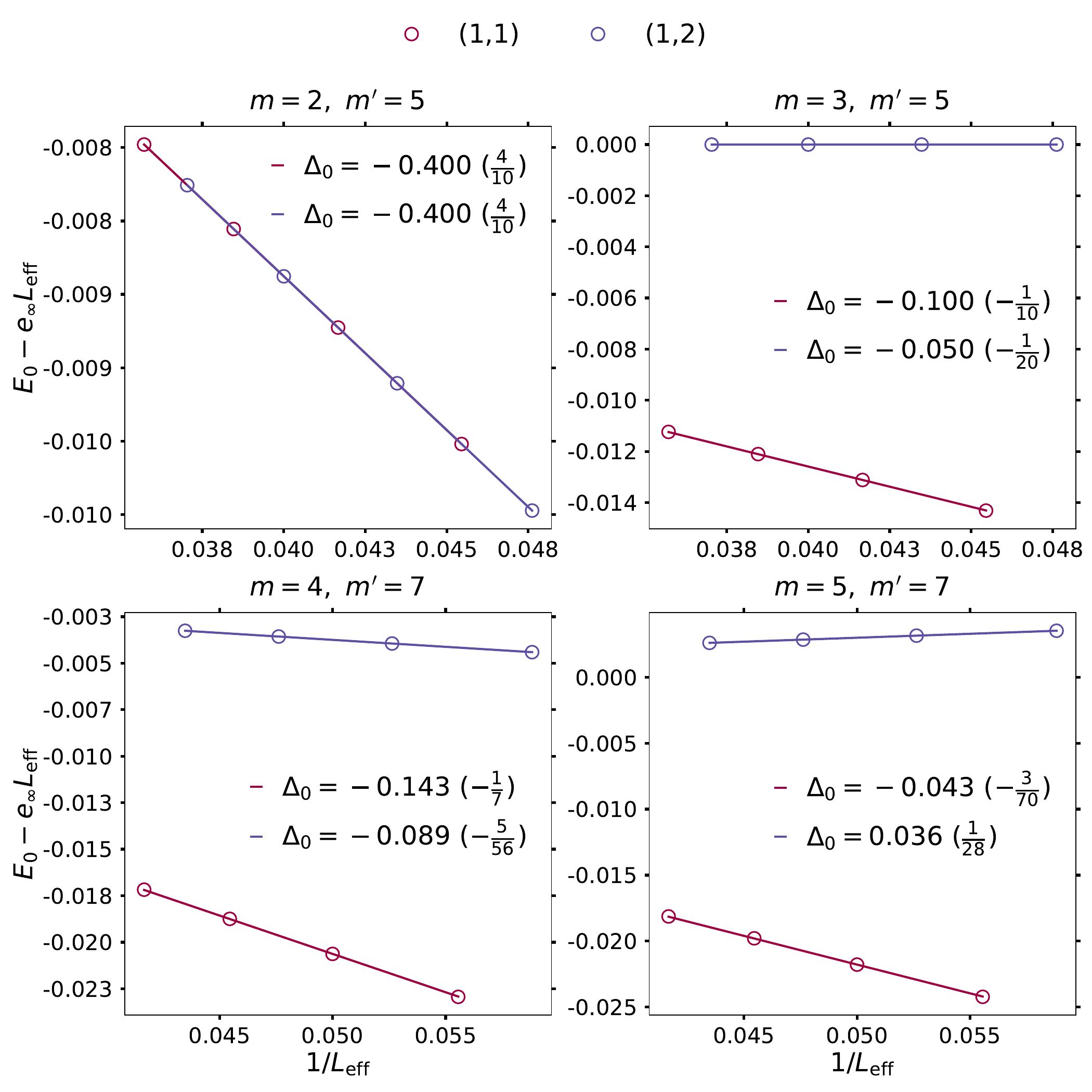}
\caption{\label{fig:fixed-point-scaling} Numerical data for the conformal dimension ($\Delta_0$) for the states with the lowest energy for
the $(1,1)$ and $(1,2)$ defects. The latter are realized by respectively setting $v$ = 0 and $v$ = $20$ in Eq.~\eqref{eq:def_Ham_simp}. The different values of $\Delta_0$ are obtained by fitting $E_0 - e_{\infty} L_{\rm eff}$ with $1/L_{\rm eff}$ [Eq. \eqref{eq:fin-sz-scal-en_1}]. Here, $e_{\infty}$ denotes the bulk
energy density given in Eq. \eqref{eq:einf-dens}. The expected CFT predictions are shown in parenthesis.}
\end{figure}
In this section, the RG fixed points corresponding to the identity~(1,1) and the KW~(1,2) defects are analyzed numerically. To that end, the low-lying eigenstates of the Hamiltonian~$H_D(v)$ are obtained for~$v = 0$ and~$v = 20$ for different RSOS$(m, m')$ models. The eigenstates are obtained by exactly diagonalizing the Hamiltonian, written as a sparse matrix in the corresponding Hilbert space built out of the states that satisfy the RSOS constraint~(see Sec.~\ref{sec:model}). For this computation, a hundred energy levels with lowest values were obtained~\footnote{For all the Hamiltonians considered in this work, the eigenvalues were obtained to be real.}. As mentioned in Sec.~\ref{sec:model}, the Hamiltonians are not Hermitian (even though they have real eigenvalues) and possess {\it distinct} left and right eigenstates. The latter are defined as:
\begin{equation}
    H \ket{\psi_R} = E_R \ket{\psi_R} \, , \quad  H^{\dagger} \ket{\psi_L} = E_L \ket{\psi_L} \,.
\end{equation}
Due to the fact that $H^{\dagger} = H^{*}$ [see below Eq.~\eqref{eq:Ham-11}], the left and right eigenvectors are related by complex conjugation.  It was numerically checked that~$E_L = E_R^* = E_R$. The expectation value of an operator~$\hat{O}$ for the given non-Hermitian system is given by \cite{brody2014biorthogonal}:
\begin{equation}
    \langle \hat{O}\rangle = \frac{\langle \psi_L|\hat{O}|\psi_R\rangle}{\langle\psi_L|\psi_R\rangle}.
\end{equation}
This was used to compute the eigenvalues of the momentum and the defect operators described in Sec.~\ref{sec:model} for a given Hamiltonian  eigenstate. 

\begin{figure}
\includegraphics[width = 0.5\textwidth]{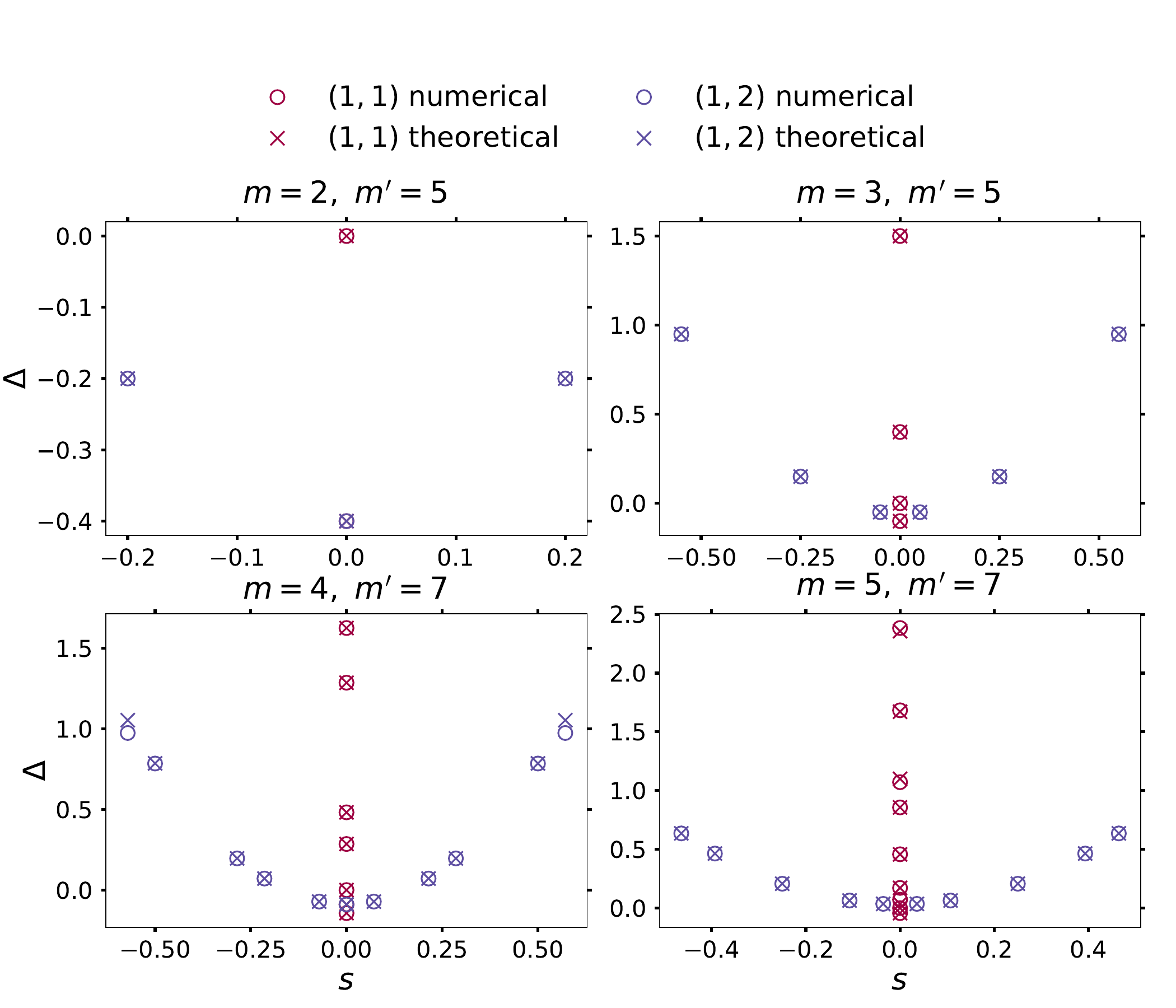}
\caption{Numerical results (in circles) for the conformal dimension $\Delta ( = h + \bar{h})$ versus the conformal spin $s( = h - \bar{h})$ for the lowest states in each of the Virasoro tower in the CFT spectrum. The CFT predictions are shown with crosses. The maroon and blue markers correspond to the (1,1) and the (1,2) defects respectively. For ${\cal M}(5,2)$ and ${\cal M}(5,3)$ all primary fields are shown, for ${ \cal M}(7,5)$ and ${\cal M}(7,4)$ some of the primary fields, which are higher up in the energy spectrum, are omitted.
}
\label{fig:deltavss}
\end{figure}

First, finite size scaling analysis is performed to identify the different eigenstates of Hamiltonian~$H_D(v)$ for the RSOS$(m,m')$ model to different states in the spectrum of the minimal~${\cal M}(m', m)$ CFT. Variation of the energy and momenta eigenvalues with system-size~$L$ yields the relevant conformal dimensions~($h, \bar{h}$). The relevant equations are given by~\cite{DiFrancesco:1997nk}:
\begin{align}
\label{eq:fin-sz-scal-en_1}
    E_{h, \bar{h}} &= e_\infty L_{\rm eff}+ \frac{2 \pi }{L_{\rm eff}}\left(-\frac{c}{12} + h + \bar{h} \right) +o\left(\frac{1}{L_{\rm eff}}\right)\, ,\\
    \label{eq:fin-sz-scal-en_2}
    P_{h, \bar{h}} &= \frac{2 \pi }{L_{\rm eff}}\left(h - \bar{h} \right) + o\left(\frac{1}{L_{\rm eff}}\right)\, .
\end{align}
Here,~$e_\infty$ is the energy density, known analytically~\cite{Koo:1993wz}
 \begin{equation}\label{eq:einf-dens}
   e_{\infty} = \frac{\gamma}{\pi}\int_{-\infty}^{\infty} dt\frac{\sinh\left(\pi - \gamma\right)t}{\sinh(\pi t)\cosh(\gamma t)} \, 
\end{equation}
and~$c$ is the central charge. Finally,~$L_{\rm eff} = L$ and $L - 1$ for the (1,1) and~(1,2) defects respectively~\cite{Grimm2001, Sinha:2023hum, Sinha:2025jhh}. 
 
Fig.~\ref{fig:fixed-point-scaling} shows the results obtained for the ground states of~$H_D(v)$ for~$v = 0$~(maroon markers) and~$v = 20$~(blue markers). This is done for four different choices of~$(m, m')$. The corresponding CFT spectra are straightforwardly inferred from Ref.~\cite{Petkova:2000ip}. The scaling dimension~$h+\bar{h}$ was obtained to be in good agreement with the theoretical predictions. Using Eqs.~(\ref{eq:fin-sz-scal-en_1}, \ref{eq:fin-sz-scal-en_2}), the conformal dimensions were identified for the primary states for the four different models (descendants are not shown for brevity). The momentum was computed for a given state by computing the corresponding expectation value of the operator~$U_k$~[Eq.~\eqref{eq:transl_op}]. This is shown in Fig.~\ref{fig:deltavss}. The numerical results for~$v = 0(20)$ are shown with maroon~(blue) circles, with the CFT predictions depicted by crosses of the same color. Note that similar results were also obtained for the~(1,1) case~\cite{Ardonne_2011} and also for the~(1,2) case, but only for the Lee-Yang model in Ref.~\cite{Lootens:2019xjv}. This work constructs the identity and KW defects for a general non-unitary minimal model, with further generalizations to other defects possible following Refs.~\cite{Sinha:2023hum, Sinha:2025jhh}. 

In addition to the analysis of the spectrum of the impurity Hamiltonian, below, numerical results are presented for the eigenvalues of the defect operator at the two fixed points. For brevity, this is done only for the~(1,2) defect operator and the primary states of the corresponding CFT. To that end, the expectation value of the transfer matrix~[Eq.~\eqref{eq:tmat_TLgen_KP}] is computed for the eigenstates of~$H_D(v = 0)$~[Eq.~\eqref{eq:def_Ham_simp}]. The choice of~$u_j = iv$ $\forall j$ with~$v = 20$ in Eq.~\eqref{eq:tmat_TLgen_KP} corresponds to the (1,2) defect. For the models under consideration, the CFT prediction for the said eigenvalue is given by~${\rm S}_{(r,s), (r', s')}/{\rm S}_{(1,1), (r',s')}$~\cite{Petkova:2000ip} where S is the corresponding modular S-matrix~\cite{DiFrancesco:1997nk} and~$(r,s)$ and~$(r', s')$ denote the defect and the primary state respectively. 

Table~\ref{tab:crossed-channel-Y-operator} shows the numerical results for the four minimal CFT models~${\cal M}(m', m)$ analyzed in Fig.~\ref{fig:deltavss}. In all cases, the corresponding numerical results were found to agree with the analytical predictions to machine precision - which is expected since the described lattice model realizes the (1,2) defect {\it exactly} on the lattice~\cite{Belletete:2018eua, Sinha:2025jhh}. Of course, corrections to scaling still appear in other observables, just not in the characteristics of certain observables for the defect. 
\begin{widetext}
\begin{table*}
\setlength{\tabcolsep}{0.5pt}
\centering
\renewcommand{\arraystretch}{1.2}
\begin{tabular}{|c|c|c|c|c|c|c|c|}
\hline
\multicolumn{2}{|c|}{${\cal M}(5,3)$} & \multicolumn{2}{c|}{${\cal M}(5,2)$} & \multicolumn{2}{c|}{${\cal M}(7,5)$} & \multicolumn{2}{c|}{${\cal M}(7,4)$} \\[6pt]
\hline
State  & $\langle Y\rangle$ & State  & $\langle Y\rangle$ & State  & $\langle Y\rangle$ & State  & $\langle Y\rangle$ \\[6pt]
\hline
$\left|-\frac{1}{20},-\frac{1}{20}\right\rangle$ & $1.618$ $(2\cos\frac{\pi}{5})$ & $\left|-\frac{1}{5},-\frac{1}{5}\right\rangle$ & $1.618$ $(2\cos\frac{\pi}{5})$ & $\left|-\frac{3}{140},-\frac{3}{140}\right\rangle$ & $1.802$ $(2\cos\frac{\pi}{7})$ & $\left|-\frac{1}{14},-\frac{1}{14}\right\rangle$ & $1.802$ $(2\cos\frac{\pi}{7})$ \\[6pt]
$\left|0,0\right\rangle$ & $0.618$ $(2\cos\frac{2\pi}{5})$ & $\left|0,0\right\rangle$ & $-0.618$ $(-2\cos\frac{2\pi}{5})$ & $\left|0,0\right\rangle$ & $1.247$ $(2\cos\frac{2\pi}{7})$ & $\left|-\frac{5}{112},-\frac{5}{112}\right\rangle$ & $1.247$ $(2\cos\frac{2\pi}{7})$ \\[6pt]
$\left|\frac{1}{5},\frac{1}{5}\right\rangle$ & $-1.618$ $(-2\cos\frac{\pi}{5})$ & & & $\left|\frac{1}{28},\frac{1}{28}\right\rangle$ & $0.445$ $(2\cos\frac{3\pi}{7})$ & $\left|0,0\right\rangle$ & $0.445$ $(2\cos\frac{3\pi}{7})$ \\[6pt]
$\left|\frac{3}{4},\frac{3}{4}\right\rangle$ & $-0.618$ $(-2\cos\frac{2\pi}{5})$ & & & $\left|\frac{3}{35},\frac{3}{35}\right\rangle$ & $-0.445$ $(-2\cos\frac{3\pi}{7})$ & $\left|\frac{1}{7},\frac{1}{7}\right\rangle$ & $-1.247$ $(-2\cos\frac{2\pi}{7})$ \\[6pt]
& & & & $\left|\frac{8}{35},\frac{8}{35}\right\rangle$ & $-1.802$ $(-2\cos\frac{\pi}{7})$ & $\left|\frac{27}{112},\frac{27}{112}\right\rangle$ & $-1.802$ $(-2\cos\frac{\pi}{7})$ \\[6pt]
& & & & $\left|\frac{3}{7},\frac{3}{7}\right\rangle$ & $-1.802$ $(-2\cos\frac{\pi}{7})$ & & \\[6pt]
\hline
\end{tabular}
\caption{Expectation value of the $Y$ operator~[Eq.~\eqref{eq:Yoperator_1}] for the states corresponding to the primary fields in non-unitary minimal models. The numerical results are obtained by constructing the corresponding operator for the lattice model with $L = 16$ sites and computing its expectation value for the primary states. The latter are identified within the eigenspectrum of~$H_D(v = 0)$ using the finite-size scaling relations~[Eqs.~(\ref{eq:fin-sz-scal-en_1}, \ref{eq:fin-sz-scal-en_2})]. The theoretical values are obtained using the S-matrix of these minimal model CFTs. Note that although we provide numerical results only up to three decimal points, they agree with the theoretical predictions up to machine precision. This is due to the fact that the corresponding defect is realized {\it exactly} for the lattice model~\cite{Belletete:2018eua, Sinha:2025jhh}.}
\label{tab:crossed-channel-Y-operator}
\end{table*}

\end{widetext}
\section{Results: Thermodynamic Entropy and RG Flows}
\label{sec:res_flows}
In the previous section, the characteristics of the different fixed points were analyzed for four different RSOS$(m, m')$ models. Now, the RG flow is analyzed between the different fixed points. For the unitary case analyzed in Refs.~\cite{Kormos:2009sk, Tavares:2024vtu}, the flow is monotonic emanating from the UV fixed point with higher g-function terminating at the IR fixed point with lower g-function~\cite{Affleck1995, Friedan2004}. For the non-unitary models analyzed in this work, a similarly monotonic RG flow is obtained. To that end, the thermodynamic entropy is computed for different choices of the impurity strength, determined by the parameter~$v$~[Eq.~\eqref{eq:def_Ham_simp}]. The latter sets the `Kondo temperature' denoted by~$1/{\rm R}_K$~\footnote{Indeed, the analyzed models can be related to non-unitary generalizations of the multi-channel Kondo models, see Ref.~\cite{Tavares:2024vtu} for discussion on the unitary case.}, where~${\rm R}_K = e^{m'v}$. As the physical temperature~(denoted by~$1/{\rm R}$) is varied,  the UV~(IR) fixed point is reached for~${\rm R}/{\rm R}_K\ll(\gg)1$. 

\begin{figure}
    \centering
    \includegraphics[width=1\linewidth]{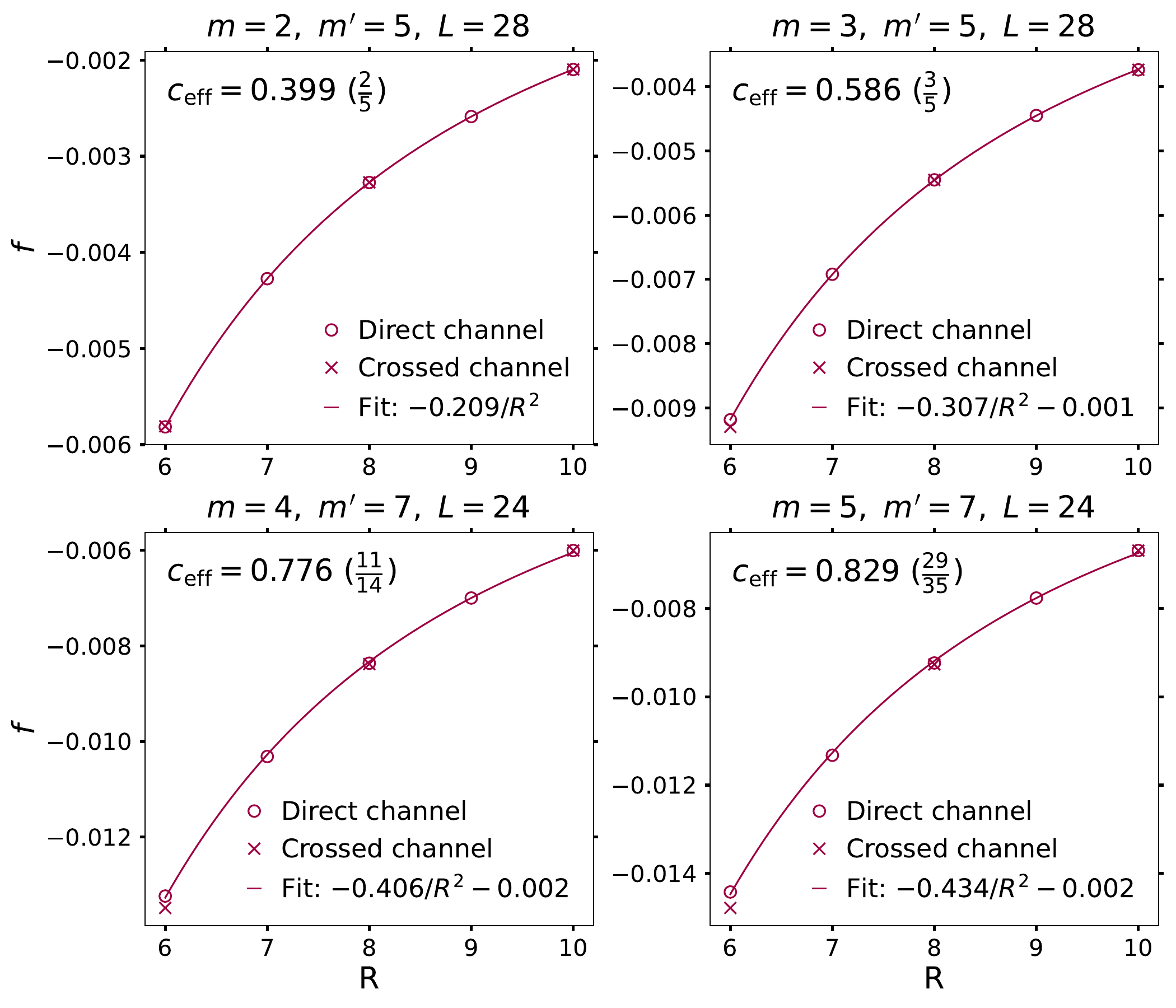}
    \caption{Free energy density in the direct channel~(denoted by circles) calculated numerically for four different RSOS$(m, m')$ models with $L = 28 (24)$ and R $= 6,7, 8, 9,10$. The effective central charge~($c_{\rm eff}$) is obtained by fitting the numerical data to Eq.~\eqref{eq:f_R_reln}. The expected results for~$c_{\rm eff}$ are shown in the parenthesis. The same free energy density is also calculated in the crossed channel~(denoted by crosses) using systems with $L = 6,8,10$ and R = 28 (24) for the corresponding  RSOS models. The agreement between the direct and crossed channel computations, as would be expected from the modular invariance of the corresponding CFT partition function, is reasonable and serves as a benchmark for the precision of our lattice computations.}
    \label{fig:f-energy-c-charge}
\end{figure}

\begin{figure}
\includegraphics[width = 0.5\textwidth]{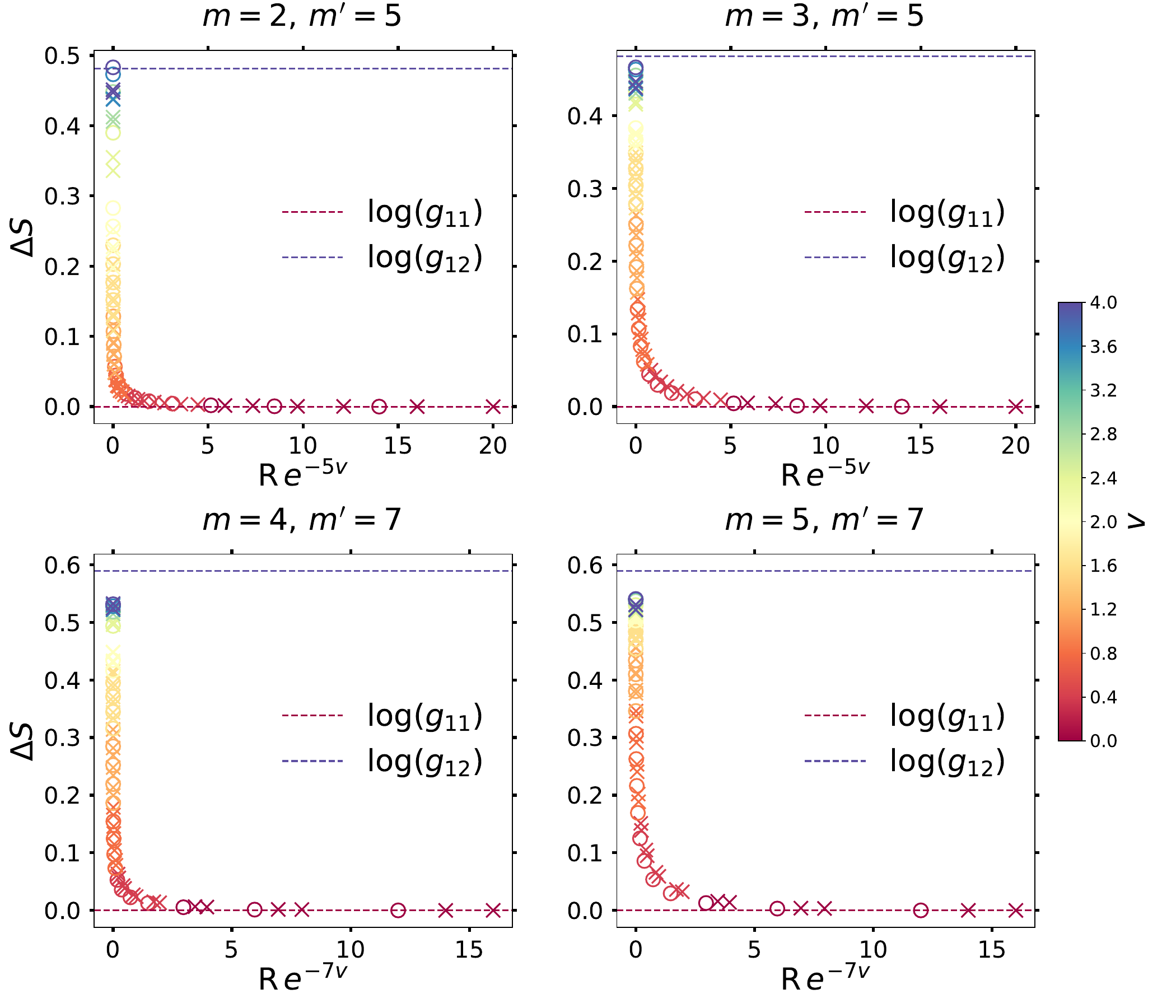}
\caption{Change in the thermodynamic entropy~$\Delta S$ upon variation of the defect parameter~$v$ for RSOS$(m,m')$ models with four different choices of~$m, m'$. As $v$ changes from 0 to 4, the lattice Hamiltonian~[Eq.~\eqref{eq:def_Ham_simp}] interpolates between the (1,1) and (1,2) fixed points. The thermodynamic entropy is computed in the direct channel~(denoted by circles) with~$L = 28 (L = 24)$ for~$m' = 5 (m' = 7)$. For this computation,~$R$ is taken to be~$L/2$. In the crossed channel, the same change~$\Delta S$ is obtained by computing the expectation value of the corresponding transfer matrix~$T(\{u\})$ with~$u_j = \gamma/2 + iv$ $\forall j$ and~$v$ varying between 0 and 4. For this computation, the system-size R was varied from 20 to 12 (16 to 10) for~$m' = 5 (m' = 7)$. As~$v$ changes,~$\Delta S$ changes from~$\ln(g_{11})$ to~$\ln(g_{12})$, where~$g_{r,s}$ is the Affleck-Ludwig g-function for defect~$(r,s)$. Here,~$g_{11} = 1$, but $g_{12} = \sin(2 \pi/5)/\sin( \pi/5)$ for $m' = 5$, and $g_{12} = \sin(2 \pi/7)/\sin( \pi/7)$ for $m' = 7$. While the direct and crossed channel computations agree to good precision, we observe discrepancy to the theoretical predictions for~$m' = 7, v\gg 1$. We believe this is due to the small size of the analyzed lattice model and expect this to improve with larger-scale numerical computations. }
\label{fig:g-func-flow}
\end{figure}
In contrast to usual Bethe ansatz or tensor network-based approaches for computation of free energy, here this is done using the exact diagonalization method. As such, the system-sizes amenable for analysis are rather small.  To benchmark our computations, first the free energy density~$f$ is computed using the lowest 100 states of $H_D(v = 0)$ for RSOS$(m, m')$ models. For~$(m, m') = (2,5)$ and (3,5), models with~$L = 28$ sites are analyzed. For ~$(m, m') = (4,7)$ and (5,7),~$L$ is taken to be 24. For all the models, the inverse temperature R is varied between 6 and 10 in steps of 1. The results are shown in Fig.~\ref{fig:f-energy-c-charge} with empty red circles and are labeled as direct channel. The obtained free energy density is fitted as a function of~$R$ to the formula~\cite{Itzykson_1986}:
\begin{equation}
    \label{eq:f_R_reln}
    f = f_0 - \frac{\pi c_{\rm eff}}{6{\rm R}^2} + \ldots,
\end{equation}
where~$f_0$ is the non-universal bulk free energy contribution and the dots indicate subleading corrections. Note that we have used the fact that for non-unitary theories, it is the effective central charge and the not the central charge that governs the behavior of the free energy. This effective central charge is given by~$c_{\rm eff} = c - 12\Delta_{\rm min}$, where~$\Delta_{\rm min}$ is the lowest scaling dimension in the corresponding CFT. In all cases, fitting with Eq.~\eqref{eq:f_R_reln} yields effective central charges that are in good agreement with the expected results. Next, the same free energy is computed by using all the eigenstates of the RSOS$(m, m')$ models of system-size R but inverse temperature~$L$, where~$(L, {\mathrm R})$ are chosen as earlier and~$v$ is kept at zero. Note that only even R-s are analyzed for this case since periodic boundary conditions are imposed. The results are shown with red crosses in Fig.~\ref{fig:f-energy-c-charge}. The agreement between the two sets of numerical data, as would be expected from modular invariance of the CFT partition functions, is reasonable. 
The precision here can potentially be further improved by including more eigenstates of~$H_D(v)$ in the direct channel computation or use approaches based on tensor networks.

\begin{figure}
\includegraphics[width = 0.5\textwidth]{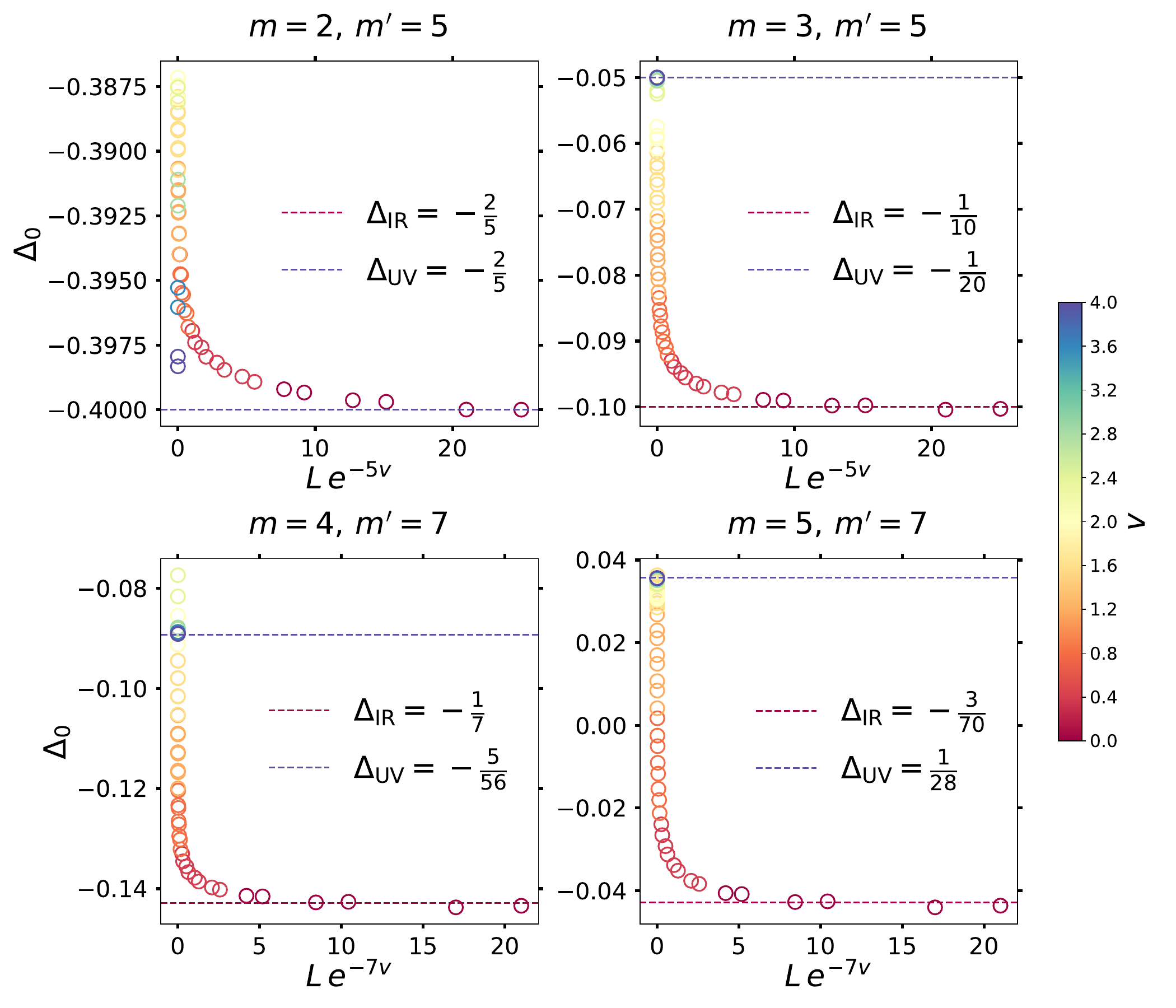}%
\caption{Change in the ground-state conformal dimension~($\Delta_0$) as the defect parameter~$v$ is varied from 0 to 4 to interpolate between the (1,1) and the (1,2) fixed points. Here,~$\Delta_0$ is calculated numerically for four RSOS$(m, m')$ models using systems of sizes $L = 18$ to $L = 28$ ($L = 14$ to $L = 24$) for $m' = 5$  $(m' = 7)$. The theoretical values for~$\Delta_0$ at the (1,1) and the (1,2) fixed points are provided and depicted using dashed horizontal lines.}
\label{fig:cd-flow}
\end{figure}

Next, the RG flow is analyzed by computing the change in the thermodynamic entropy~($\Delta S$) for the four models by considering different values of~$v$. For a given choice of~$L$ and R, the thermodynamic entropy in the direct channel by considering the lowest 100 eigenstates of~$H_D(v)$ for~$v$ varying between 0 and 4.0. Fig.~\ref{fig:g-func-flow} shows the results of this computation for the same choices of~$L$ and~$(m, m')$ as in Fig.~\ref{fig:f-energy-c-charge}. The change in the thermodynamic entropy, computed with respect to that obtained for~$v = 0$, captures the contribution from the defect. As~$R/{\rm R}_K$ is varied,~$\Delta S$ varies from 0 to~$\ln(g_{12}/g_{11})$. The results obtained for different choices of R and $v$, shown with circles of different colors, collapse on to a single scaling curve, indicative of the fact that the field theoretic regime is indeed being probed using the lattice model. The change in thermodynamic entropy can also be calculated using Thermodynamic Bethe Ansatz, as was done for RSOS$(k,k+1)$, where $k = 3,4,5$, in Ref.~\cite{Tavares:2024vtu}. In this paper, we restrict ourselves to calculation of the defect entropy using ab-initio lattice calculations. 

The change in the defect entropy is also computed in the crossed channel. This is done by computing the expectation value of the transfer matrix~$T(\{u\})$~[Eq.~\eqref{eq:tmat_TLgen_KP}] for R sites. Here,~$u_j$ is taken to be $\gamma/2 + iv$ $\forall j$ and~$v$ ranges between 0 and 4.0. The subleading~${\cal O}(1)$ term in~$\ln\langle T(\{u\})\rangle$ yields~$\Delta S$~[see Ref.~\cite{Tavares:2024vtu} for details on how to subtract the leading term in the expression for~$\ln\langle T(\{u\})\rangle$]. The results are shown with crosses of different colors in Fig.~\ref{fig:g-func-flow}. As expected, the crossed channel results collapse on to the curve obtained earlier. Note that for~${\rm R}/{\rm R}_K\ll1$, the agreement with the theoretical prediction is not as well as for~${\rm R}/{\rm R}_K\gg1$. We believe this is a finite-size effect and the agreement can be further improved by increasing the $L$ and R in the computations. 

Before concluding, we note that while the RG flow is well-captured using the thermodynamic entropy, a different signature of the same flow can be obtained by considering the system at zero temperature. In this scenario, the defect parameter~$v$ sets the `screening length'~$\xi$ given by~$e^{m'v}$. The flow between the two fixed points is obtained by varying the ratio~$L/\xi$. Fig.~\ref{fig:cd-flow} shows the change in the scaling dimension~($h+\bar{h}$) for the ground state of the impurity Hamiltonian~$H_D(v)$~[Eq.~\eqref{eq:def_Ham_simp}] for four different choices of~$m', m$. At the fixed points, the numerical results are in good agreement with the theoretical predictions obtained from CFT partition function.

\section{Conclusion}\label{sec:concl}
To summarize, in this work, topological defects are analyzed in non-unitary minimal CFTs. This is done using non-unitary variants of quantum chains derived from RSOS models. Signatures of the topological defects at the fixed points as well as RG flows connecting the latter are obtained from the eigenspectra of the corresponding lattice Hamiltonians. The change in the thermodynamic entropy is computed along the RG flow connecting the two fixed points. While realizations are provided explicitly for the identity and the KW defects, similar computations can potentially be done for other defects generalizing the approaches of Refs.~\cite{Sinha:2023hum, Sinha:2025jhh}. Note that even though the computations performed in this work are numerical, due to the integrable nature of the lattice models, the corresponding quantities can also be obtained using Bethe ansatz techniques. Finally, generalizations of these models using the XXZ representation of the TL algebra lead to non-Hermitian Kondo-type models~\cite{Nakagawa2018, Yoshimura2020}. The latter, while close to the analyzed RSOS family of models, are interesting on their own. We leave the analysis of these models for a future work.

\begin{acknowledgments}
The authors acknowledge discussions with Gleb Kotousov, Sergei Lukyanov, David Rogerson, Sahand Seifnashri, and Shu-Heng Shao. AR was supported by the U.S. Department of Energy, Office of Basic Energy Sciences, under Contract No. DE-SC0012704. 
\end{acknowledgments}

\appendix

\begin{widetext}

\section{Anyonic chain construction}\label{sec:any-chain}

Anyonic chains are 1+1D lattice models, defined using $F$-matrices of fusion categories \cite{Feiguin:2006ydp}. In many cases, these lattice models flow to 1+1D CFT. For instance, in the case of $\mathsf{su}(2)_k$ fusion category \cite{Gils_2009, Aasen:2020jwb}, these lattice models \footnote{upon choice of certain simple object and Hamiltonian} are equivalent to RSOS$(k+1,k+2)$ models, and flow to the unitary ${\cal M}(k+2, k+1)$ minimal model CFT \cite{Sinha:2025jhh}. In \cite{Ardonne_2011, Lootens:2019xjv}, it was shown that non-unitary minimal model CFTs can be obtained using Galois conjugated $\mathsf{su(2)}_{k}$ categories. In this appendix, we show that the anyonic chain lattice models of \cite{Ardonne_2011,Lootens:2019xjv} are equivalent to the RSOS models we discussed in this paper.

Let us recall the data of the category $\mathsf{su(2)}_{k}$  first. Its simple objects are $1, 2,\ldots,k+1$ and the fusion rule are given by 
\begin{equation}\label{eq:fus-rul-su-A}
    N^{j_1}_{j_2 j_3} = 
    \begin{cases} 
        1 \quad 
        \begin{aligned}
         & \text{if }    j_1 + j_2 \geq j_3 + 1 , \quad j_2 + j_3 \geq j_1 + 1 , \quad j_3 + j_1 \geq j_2 + 1 \, ,  \\ 
            & j_1 + j_2 + j_3 \in 2 \mathbb{Z} + 1  , \quad j_1 + j_2 + j_3 \leq 2k+3 \, , 
        \end{aligned}  \\ 
        0 \quad \text{otherwise.}
    \end{cases}
\end{equation}
To define the $F-$symbols for the unitary categories, we define the following 
\begin{equation}\label{eq:params}
\begin{split}
    q & = {\rm e}^{\frac{{\rm i } \pi}{k+2}} \, ,  \\
    [ m] & = \frac{q^{m} - q^{-m}}{q -q^{-1}} \, ,  \\
    [n]! &=
\begin{cases} 
\prod_{m=1}^n [m] & n > 0, \\
1 & n = 0, \\
\infty & n < 0,
\end{cases} \\
\Delta(j_1, j_2, j_3) &=
\begin{cases} 
\sqrt{   \frac{\left[\frac{j_1+j_2-j_3 - 1}{2} \right]!\left[\frac{j_3+j_1-j_2 - 1}{2}\right]!\left[\frac{j_2+j_3-j_1-1}{2}\right]!}{\left[ \frac{j_1+j_2+j_3-1}{2}\right]!}} & N_{j_2j_3}^{j_1} = 1, \\
0 & \text{else}.
\end{cases}
\end{split}
\end{equation}
using which we define the $q-$deformed Wigner 6j symbols 
 \cite{Kirillov:1991ec}
\begin{equation}\label{eq:qdef-Wig}
\begin{split}
& \left\{\begin{array}{lll}
j_1 & j_2 & j_3 \\
j_4 & j_5 & j_6
\end{array}\right\}_q=\Delta\left(j_1, j_2, j_3 \right) \Delta\left(j_1, j_5, j_6\right) \Delta\left(j_5, j_3, j_4\right) \Delta\left(j_4, j_2, j_6\right) \\
& \times \sum_z  \frac{(-1)^z[z-1]!}{\left[z - \frac{j_1+j_2+j_3 + 1}{2}\right]!\left[z- \frac{j_1 + j_5 + j_6 + 1}{2} \right]!\left[z- \frac{j_2 + j_4 + j_6 + 1}{2}  \right]!\left[z- \frac{j_4+j_5+j_3 + 1}{2}    \right]!}
\\
& \times \quad \frac{1}{\left[ \frac{j_1 + j_2 + j_4 + j_5}{2} - z \right]!\left[ \frac{j_1+j_4+j_3+j_6}{2}-z\right]!\left[\frac{j_2+j_5+j_3+j_6}{2}-z\right]!}\, .
\end{split}
\end{equation} 
     Here, $z$ is an integer with the constraint that the terms in the summation are finite.
We use these to define the $\mathsf{su(2)}_{k}$ $F$ symbol 
\begin{equation}\label{eq:Fsymb-su2}
    \left({F}^{abc}_d\right)_{e,f} = (-1)^{\frac{a+b+c+d}{2}} \sqrt{[e]} \sqrt{  [f]}
\begin{Bmatrix}
a & b & e \\
c & d & f
\end{Bmatrix}_q
 \, .
\end{equation}

Note, for a simple object $a$
\begin{equation}
    d_a = [a] =\frac{\sin \left( \frac{\pi a}{k+2} \right)}{\sin \left( \frac{\pi}{k+2} \right)}\, , 
\end{equation}
is called quantum dimension of $a$. The quantum dimensions are positive real numbers for  ${\mathsf su}(2)_{k}$. 

\bigskip

Using these categories, we can define lattice models called anyonic chains \cite{Feiguin:2006ydp}. The Hilbert space for the anyonic chain with $L$ sites is defined using the fusion tree of fig. \ref{fig:fusiontree}.
\begin{figure}[h]
\includegraphics[width=0.5\linewidth]{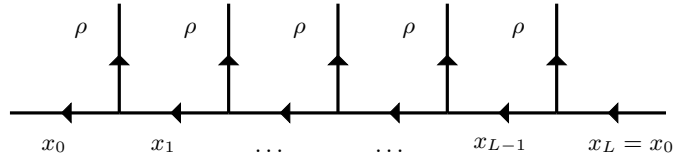}
\caption{The fusion tree that defines the anyonic chain. In this work we set $\rho = 2$ }
\label{fig:fusiontree}
\end{figure}
It is easy to see that the Hilbert space of RSOS$(\cdot,k+2)$ model is the same as that of  $\mathsf{su}(2)_{k}$, when we set $\rho = 2$. 
 The Hamiltonian for the anyonic chain is given by \cite{Buican:2017rxc} 
\begin{equation}\label{eq:Ham-anyonic}
    H^{a,\rho} = - \sum_{j}  J\,  h^{a,\rho}_j \, ,
\end{equation}
where
\begin{equation}\label{eq:Ham-all-spin}
\begin{split}    
      & h_{j}^{a,\rho} \ket{\ldots x_{j-1}, x_j, x_{j + 1} \ldots }   \\ & = \sum_{ \tilde{x}_j  } \,    \left(F^{x_{j-1} \,  \rho \, \rho}_{x_{j+1}}\right)_{x_j,  a} \, \,  \left(F^{-1 \, x_{j-1} \, \rho \, \rho}_{x_{j+1}} \right)_{a,  \tilde{x}_j}   \ket{\ldots x_{j-1}, \tilde{x}_j, x_{j+1} \ldots } \, ,
\end{split}
\end{equation}
Note, here $a$ is the label of the anyon in whose channel we project in to. From now, we only consider the case $\rho = 2$ and $a = 1$, then it can be checked that for $h_j^{1, 2}$ to be non-zero, $x_{j-1} = x_{j+1}$ and $\wt{x}_j = x_{j-1} \pm 1$. Hence, upon re-writing Eq.~\eqref{eq:Ham-all-spin} for $\rho = 2$ and $a = 1$ we get 
\begin{equation}\label{eq:spin-1/2-ham}
\begin{split}
&        h_{j}^{1 ,2} \ket{\ldots x_{j-1}, x_j, x_{j + 1} \ldots } = \\ & \delta_{x_{j-1}, x_{j+1}}  \sum_{ \tilde{x}_j \in \{ x_{j-1} - 1,x_{j-1} +  1\}  } \,    \left(F^{x_{j-1} \,  2 \,  2 }_{x_{j-1}}\right)_{x_j, \,  1} \, \,  \left(F^{x_{j-1} \, 2  \, 2 }_{x_{j-1}} \right)_{\tilde{x}_j, \, 1}   \ket{\ldots x_{j-1}, \tilde{x}_j, x_{j+1} \ldots } \, ,
\end{split}
\end{equation}
where we have used the fact that $\left(F^{x_{j-1} \, 2  \, 2 }_{x_{j-1}} \right)^{T} = \left(F^{-1 \, x_{j-1} \, 2  \, 2 }_{x_{j-1}} \right) $. Using the $F-$symbols of $\mathsf{su}(2)_k $, it was shown in \cite{Sinha:2025jhh} that 

\begin{equation}\label{eq:h-gen-two-F}
\begin{split}    
 &\bra{\ldots x_{j-1}, \tilde{x}_j, x_{j+1}, \ldots}   {h}^{1,2}_j\ket{\ldots x_{j-1}, x_j, x_{j+1}, \ldots} = \delta_{x_{j-1},x_{j+1}} (-1)^{x_{j} - \tilde{x}_j} \frac{\sqrt{d_{x_j}}\sqrt{d_{\tilde{x}_j}}}{d_{2} \,  d_{x_{j-1}}} \, , \\ 
 \end{split}
\end{equation}
Consider the following unitary operator 
\begin{equation}
    T_h \ket{x_0, x_1, \ldots, x_{L - 1}}   = (-1)^{\sum \frac{x_i}{2}}\ket{x_0, x_1, \ldots, x_{L - 1}} \, , 
\end{equation}
then it is easy to see that  
\begin{equation}\label{eq:T_h_op_su2k}
    e_i = d_{2}T_h^{\dagger} \, \hat{h}_i   \, T_h \, , \\
\end{equation}
where $e_i$ is the TL generator for RSOS$(k+1,k+2)$. As $h_j^{1,2}$ are symmetric and real, it is easy to see that  $H^{1,2}$ is Hermitian. 

\bigskip

Non-Hermitian Hamiltonians can be obtained by a process called Galois conjugation, we briefly describe this here. The $F-$symbols given in Eq.~\eqref{eq:Fsymb-su2} are solutions to certain consistency conditions called the Pentagon equations. Fixing the fusion rules, the Ocneanu rigidity theorem states that there are only finitely many inequivalent solutions \footnote{not counting the ones differing by gauge transformations} to the Pentagon relations \cite{etingof2005fusion}. The Galois group relates the different inequivalent solutions and Galois conjugation is the way of moving between these different solutions. The Galois conjugated MTCs may be non-unitary as well. See \cite{buican2022galois} for a nice review on the importance of Galois conjugation in the study of topological QFTs. By deforming $q$, using an integer $p$ in the following way 
\begin{equation}
    q(p) = {\rm e}^{\frac{{\rm i} p \pi}{k+2}} \, ,
\end{equation}
where $2 \leq p \leq \frac{k+2}{2}$ and $p$ and $k+2$ are co-prime, we obtain Galois conjugated MTCs for  $\mathsf{su(2)}_{k}$. The $F-$symbol for this category is obtained by plugging $q(p)$ into Eq.~\eqref{eq:params}, \eqref{eq:qdef-Wig}, and \eqref{eq:Fsymb-su2}.
The new quantum dimension for simple object $a$ is given by 
\begin{equation}
    d(p)_a = \frac{q(p)^{a} - q(p)^{-a}}{q(p) -q(p)^{-1}}  = \frac{\sin\left(\frac{ a p \pi}{k+2} \right)}{\sin \left(\frac{p \pi}{k+2} \right)} \, .
\end{equation}

Note, while the quantum dimensions are always positive in the case of unitary $p = 1$  ${\mathsf su}(2)_{k}$ category, they can be negative in the Galois conjugated case.

\bigskip 

Let us now discuss the anyonic chain model constructed using Galois conjugated category. As the fusion rules do not change, the Hilbert space remains the same.  Similar to ${h}^{1,2}_j$, we can define ${h}(p)^{1,2}_j$
\begin{equation}
\begin{split}
&  {h}(p)^{1,2}_j\ket{\ldots x_{j-1}, x_j, x_{j+1}, \ldots}  \\
 = \,  &  \delta_{x_{j-1}, x_{j+1}}  \sum_{ \tilde{x}_j \in \{ x_{j-1} -1,x_{j-1} +  1\}  } \,    _p\left({F}^{x_{j-1} \,  2 \,  2 }_{x_{j-1}}\right)_{x_j, \,  1} \, \,  _p\left({F}^{-1  \, x_{j-1} \, 2  \, 2 }_{x_{j-1}} \right)_{ 1 , \, \tilde{x}_j  }   \ket{\ldots x_{j-1}, \tilde{x}_j, x_{j+1} \ldots }  \, , \\
 = \,  &  \delta_{x_{j-1}, x_{j+1}}  \sum_{ \tilde{x}_j \in \{ x_{j-1} - 1,x_{j-1} +  1\}  } \,    _p\left({F}^{x_{j-1} \,  2 \,  2 }_{x_{j-1}}\right)_{x_j, \,  1} \, \,  _p\left({F}^{   x_{j-1} \, 2  \, 2}_{x_{j-1}} \right)_{ \tilde{x}_j , \,  1  }   \ket{\ldots x_{j-1}, \tilde{x}_j, x_{j+1} \ldots }  \, , 
 \end{split}
\end{equation}
where the second equality above follows as $_p\left(F^{x_{j-1} \, 2  \, 2 }_{x_{j-1}} \right)^{T} = \, _p\left(F^{-1 \, x_{j-1} \, 2  \, 2 }_{x_{j-1}} \right)$.  Then by inputting the Galois conjugated $F-$symbol above, it was checked numerically for $\mathsf{su(2)}$ $k = 3, 5$ that 
\begin{equation}
    \begin{split}
      & \bra{\ldots , x'_{j-1}, x'_j, x'_{j+1}, \ldots}{h}(p)^{1,2}_j \ket{\ldots, x_{j-1},x_{j},x_{j+1},\ldots} =  \\
      & (-1)^{x_j - x'_j}\left(\prod_{i \neq j } \delta_{x'_i,  x_i} \right) \delta_{x_{j-1}, x_{j+1}}    \frac{ \sqrt{ d(p)_{x_j } } \sqrt{ d(p)_{x'_j } } } { d(p)_{2} \, d(p)_{x_{j-1}}}
      \, . 
\end{split}
\end{equation}
The above defined operator is still symmetric, but as the quantum dimensions are allowed to be negative in the Galois-conjugated case, this operator is not real. Thus the Hamiltonian
\begin{equation}
        H^{1,2}(p) = - \sum_{j}  J  \, h(p)^{1,2}_j \, ,
\end{equation}
is non-Hermitian in general. Now, recall the function we defined in Eq.~\eqref{eq:S-mat-WZW}, if we set $m' = k + 2$,  we can re-write the above eqs. as 
\begin{equation}
    \begin{split}
      & \bra{\ldots , x'_{j-1}, x'_j, x'_{j+1}, \ldots}h(p)^{1,2}_j \ket{\ldots, x_{j-1},x_{j},x_{j+1},\ldots} =  \\
      & (-1)^{x_j - x'_j}\left(\prod_{i \neq j } \delta_{x'_i,  x_i} \right) \delta_{x_{j-1}, x_{j+1}}    \frac{ \sqrt{ S_{p, x_j} } \sqrt{ S_{p, x'_j }} } { d(p)_{2} \,  S_{p,x_{j - 1} } }
      \, . 
\end{split}
\end{equation}
Comparing the above equation with Eq.~\eqref{eq:TL-op}, we see that 
\begin{equation}
         h(p)^{1,2}_j = d(p)_{2} \,  T_h^{\dagger}  e_j T_h \, , 
\end{equation}
where $e_j$ is the TL generator of RSOS$(k+2-p,k+2)$.

\bigskip

 The claim in \cite{Ardonne_2011} is that when $J > 0$ in Eq.~\eqref{eq:Ham-anyonic}, i.e. anti-ferromagnetic coupling,  and $1 \leq p \leq \frac{k+2}{2} $ and $k+2$ are co-prime, then the anyonic chain model flows to the ${\cal M }(k+2,k+2 - p)$ minimal model CFT.  But as we discussed above, this implies that for $1\leq p \leq \frac{k+2}{2} $, RSOS$(k+2 - p, k + 2)$ flows to ${\cal M }(k+2,k+2 - p)$ minimal model CFT.

\bigskip 

 Further, \cite{Ardonne_2011} claims that for $J < 0$ in Eq.~\eqref{eq:Ham-anyonic}, i.e. ferromagnetic coupling, we flow to ${\cal M }( k+2,p)$ minimal model CFT, again with $2\leq p \leq \frac{k+2}{2} $ and $k+2$  co-prime. So RSOS$(k+2 - p, k + 2)$ with ferromagnetic coupling flows to ${\cal M }( p,k+2)$. But as we noted before, RSOS$(k+2 - p, k + 2)$ with ferromagnetic coupling is equivalent to RSOS$(p,k + 2)$ with anti-ferromagnetic coupling, which therefore flows to ${\cal M }(k+2, p)$. 

\bigskip
Combining the two above observations, we can simply summarize that for
$2\leq k' \leq k+1 $, RSOS$(k', k + 2)$ flows to ${\cal M }(k+2,k')$ minimal model CFT, when $k'$ and $k + 2$ are co-prime.

\end{widetext}

\bibliography{library_1}

@article{Nakayama:2024msv,
    author = "Nakayama, Yu and Tanaka, Takahilo",
    title = "{Infinitely many new renormalization group flows between Virasoro minimal models from non-invertible symmetries}",
    eprint = "2407.21353",
    archivePrefix = "arXiv",
    primaryClass = "hep-th",
    reportNumber = "YITP-24-92",
    doi = "10.1007/JHEP11(2024)137",
    journal = "JHEP",
    volume = "11",
    pages = "137",
    year = "2024"
}

@article{Lamb:2025rbl,
  title={Signatures of Topological Symmetries on a Noisy Quantum Simulator},
  author={Lamb, Christopher and Konik, Robert M and Saleur, Hubert and Roy, Ananda},
  journal={arXiv preprint arXiv:2510.14817},
  year={2025}
}

@article{Dijkgraaf:1991ba,
    author = "Dijkgraaf, Robbert and Verlinde, Herman L. and Verlinde, Erik P.",
    title = "{String propagation in a black hole geometry}",
    reportNumber = "PUPT-1252, IASSNS-HEP-91-22",
    doi = "10.1016/0550-3213(92)90237-6",
    journal = "Nucl. Phys. B",
    volume = "371",
    pages = "269--314",
    year = "1992"
}

@article{Maldacena:2000kv,
    author = "Maldacena, Juan Martin and Ooguri, Hirosi and Son, John",
    title = "{Strings in AdS(3) and the SL(2,R) WZW model. Part 2. Euclidean black hole}",
    eprint = "hep-th/0005183",
    archivePrefix = "arXiv",
    reportNumber = "CALT-68-2266, CITUSC-00-021, HUTP-00-A009, UCB-PTH-00-10",
    doi = "10.1063/1.1377039",
    journal = "J. Math. Phys.",
    volume = "42",
    pages = "2961--2977",
    year = "2001"
}

@article{Nakagawa2018,
  title = {Non-Hermitian Kondo Effect in Ultracold Alkaline-Earth Atoms},
  author = {Nakagawa, Masaya and Kawakami, Norio and Ueda, Masahito},
  journal = {Phys. Rev. Lett.},
  volume = {121},
  issue = {20},
  pages = {203001},
  numpages = {6},
  year = {2018},
  month = {Nov},
  publisher = {American Physical Society},
  doi = {10.1103/PhysRevLett.121.203001},
  url = {https://link.aps.org/doi/10.1103/PhysRevLett.121.203001}
}

@article{Bazhanov:2020uju,
    author = "Bazhanov, Vladimir V. and Kotousov, Gleb A. and Lukyanov, Sergei L.",
    title = "{Equilibrium density matrices for the 2D black hole sigma models from an integrable spin chain}",
    eprint = "2010.10603",
    archivePrefix = "arXiv",
    primaryClass = "hep-th",
    reportNumber = "DESY 20-165, DESY-20-165",
    doi = "10.1007/JHEP03(2021)169",
    journal = "JHEP",
    volume = "03",
    pages = "169",
    year = "2021"
}

@article{Hatano1996,
  title = {Localization Transitions in Non-Hermitian Quantum Mechanics},
  author = {Hatano, Naomichi and Nelson, David R.},
  journal = {Phys. Rev. Lett.},
  volume = {77},
  issue = {3},
  pages = {570--573},
  numpages = {0},
  year = {1996},
  month = {Jul},
  publisher = {American Physical Society},
  doi = {10.1103/PhysRevLett.77.570},
  url = {https://link.aps.org/doi/10.1103/PhysRevLett.77.570}
}

@article{Shen2018,
  title = {Topological Band Theory for Non-Hermitian Hamiltonians},
  author = {Shen, Huitao and Zhen, Bo and Fu, Liang},
  journal = {Phys. Rev. Lett.},
  volume = {120},
  issue = {14},
  pages = {146402},
  numpages = {6},
  year = {2018},
  month = {Apr},
  publisher = {American Physical Society},
  doi = {10.1103/PhysRevLett.120.146402},
  url = {https://link.aps.org/doi/10.1103/PhysRevLett.120.146402}
}

@article{Altland1997,
  title = {Nonstandard symmetry classes in mesoscopic normal-superconducting hybrid structures},
  author = {Altland, Alexander and Zirnbauer, Martin R.},
  journal = {Phys. Rev. B},
  volume = {55},
  issue = {2},
  pages = {1142--1161},
  numpages = {0},
  year = {1997},
  month = {Jan},
  publisher = {American Physical Society},
  doi = {10.1103/PhysRevB.55.1142},
  url = {https://link.aps.org/doi/10.1103/PhysRevB.55.1142}
}

@article{Ludwig_2016,
doi = {10.1088/0031-8949/2015/T168/014001},
url = {https://doi.org/10.1088/0031-8949/2015/T168/014001},
year = {2015},
month = {dec},
publisher = {IOP Publishing},
volume = {2016},
number = {T168},
pages = {014001},
author = {Ludwig, Andreas W W},
title = {Topological phases: classification of topological insulators and superconductors of non-interacting fermions, and beyond},
journal = {Physica Scripta}
}

@article{miri2019exceptional,
  title={Exceptional points in optics and photonics},
  author={Miri, Mohammad-Ali and Alu, Andrea},
  journal={Science},
  volume={363},
  number={6422},
  pages={eaar7709},
  year={2019},
  publisher={American Association for the Advancement of Science}
}

@article{weidemann2020topological,
  title={Topological funneling of light},
  author={Weidemann, Sebastian and Kremer, Mark and Helbig, Tobias and Hofmann, Tobias and Stegmaier, Alexander and Greiter, Martin and Thomale, Ronny and Szameit, Alexander},
  journal={Science},
  volume={368},
  number={6488},
  pages={311--314},
  year={2020},
  publisher={American Association for the Advancement of Science}
}

@article{Wang:2020gqg,
    author = "Wang, Kai and Dutt, Avik and Yang, Ki Youl and Wojcik, Casey C. and Vu{\v{c}}kovi{\'c}, Jelena and Fan, Shanhui",
    title = "{Observation of arbitrary topological windings of a non-Hermitian band}",
    eprint = "2011.14275",
    archivePrefix = "arXiv",
    primaryClass = "physics.optics",
    doi = "10.1126/science.abf6568",
    journal = "Science",
    volume = "371",
    pages = "1240",
    year = "2021"
}

@article{ding2022non,
  title={Non-Hermitian topology and exceptional-point geometries},
  author={Ding, Kun and Fang, Chen and Ma, Guancong},
  journal={Nature Reviews Physics},
  volume={4},
  number={12},
  pages={745--760},
  year={2022},
  publisher={Nature Publishing Group UK London}
}

@article{Kawabata2019,
  title = {Symmetry and Topology in Non-Hermitian Physics},
  author = {Kawabata, Kohei and Shiozaki, Ken and Ueda, Masahito and Sato, Masatoshi},
  journal = {Phys. Rev. X},
  volume = {9},
  issue = {4},
  pages = {041015},
  numpages = {52},
  year = {2019},
  month = {Oct},
  publisher = {American Physical Society},
  doi = {10.1103/PhysRevX.9.041015},
  url = {https://link.aps.org/doi/10.1103/PhysRevX.9.041015}
}

@article{Zhou2019,
  title = {Periodic table for topological bands with non-Hermitian symmetries},
  author = {Zhou, Hengyun and Lee, Jong Yeon},
  journal = {Phys. Rev. B},
  volume = {99},
  issue = {23},
  pages = {235112},
  numpages = {19},
  year = {2019},
  month = {Jun},
  publisher = {American Physical Society},
  doi = {10.1103/PhysRevB.99.235112},
  url = {https://link.aps.org/doi/10.1103/PhysRevB.99.235112}
}

@article{Gong2018,
  title = {Topological Phases of Non-Hermitian Systems},
  author = {Gong, Zongping and Ashida, Yuto and Kawabata, Kohei and Takasan, Kazuaki and Higashikawa, Sho and Ueda, Masahito},
  journal = {Phys. Rev. X},
  volume = {8},
  issue = {3},
  pages = {031079},
  numpages = {33},
  year = {2018},
  month = {Sep},
  publisher = {American Physical Society},
  doi = {10.1103/PhysRevX.8.031079},
  url = {https://link.aps.org/doi/10.1103/PhysRevX.8.031079}
}

@article{Bergholtz2021,
  title = {Exceptional topology of non-Hermitian systems},
  author = {Bergholtz, Emil J. and Budich, Jan Carl and Kunst, Flore K.},
  journal = {Rev. Mod. Phys.},
  volume = {93},
  issue = {1},
  pages = {015005},
  numpages = {31},
  year = {2021},
  month = {Feb},
  publisher = {American Physical Society},
  doi = {10.1103/RevModPhys.93.015005},
  url = {https://link.aps.org/doi/10.1103/RevModPhys.93.015005}
}

@article{Sinha:2025jhh,
  title={Integrability and lattice discretizations of all Topological Defect Lines in minimal CFTs},
  author={Sinha, Madhav and Tavares, Thiago Silva and Roy, Ananda and Saleur, Hubert},
  journal={Nuclear Physics B},
  pages={117308},
  year={2026},
  publisher={Elsevier}
}

@article{Gaiotto:2014kfa,
	author = "Gaiotto, Davide and Kapustin, Anton and Seiberg, Nathan and Willett, Brian",
	title = "{Generalized Global Symmetries}",
	eprint = "1412.5148",
	archivePrefix = "arXiv",
	primaryClass = "hep-th",
	doi = "10.1007/JHEP02(2015)172",
	journal = "JHEP",
	volume = "02",
	pages = "172",
	year = "2015"
}

@article{Roy:2024xdi,
    author = "Roy, Ananda",
    title = "{Variational Quantum Simulation of Anyonic Chains}",
    journal = "arXiv:2412.17781",
    archivePrefix = "arXiv",
    primaryClass = "quant-ph",
    month = "12",
    year = "2024"
}

@article{Roy2024,
  title = {Topological interfaces of Luttinger liquids},
  author = {Roy, Ananda and Saleur, Hubert},
  journal = {Phys. Rev. B},
  volume = {109},
  issue = {16},
  pages = {L161107},
  numpages = {6},
  year = {2024},
  month = {Apr},
  publisher = {American Physical Society},
  doi = {10.1103/PhysRevB.109.L161107},
  url = {https://link.aps.org/doi/10.1103/PhysRevB.109.L161107}
}

@article{Tavares:2024vtu,
    author = "Tavares, Thiago Silva and Sinha, Madhav and Grans-Samuelsson, Linnea and Roy, Ananda and Saleur, Hubert",
    title = "{Integrable RG Flows on Topological Defect Lines in 2D Conformal Field Theories}",
    journal = "arXiv:2408.08241",
    eprint = "2408.08241",
    archivePrefix = "arXiv",
    primaryClass = "hep-th",
    month = "8",
    year = "2024"
}

@article{Friedan2004,
  title = {Boundary Entropy of One-Dimensional Quantum Systems at Low Temperature},
  author = {Friedan, Daniel and Konechny, Anatoly},
  journal = {Phys. Rev. Lett.},
  volume = {93},
  issue = {3},
  pages = {030402},
  numpages = {4},
  year = {2004},
  month = {Jul},
  publisher = {American Physical Society},
  doi = {10.1103/PhysRevLett.93.030402},
  url = {https://link.aps.org/doi/10.1103/PhysRevLett.93.030402}
}

@article{Schafer-Nameki:2023jdn,
    author = "Schafer-Nameki, Sakura",
    title = "{ICTP lectures on (non-)invertible generalized symmetries}",
    eprint = "2305.18296",
    archivePrefix = "arXiv",
    primaryClass = "hep-th",
    doi = "10.1016/j.physrep.2024.01.007",
    journal = "Phys. Rept.",
    volume = "1063",
    pages = "1--55",
    year = "2024"
}

@article{Katsevich:2025ojk,
  title={Towards a Quintic Ginzburg-Landau Description of the (2, 7) Minimal Model},
  author={Katsevich, Andrei and Klebanov, Igor and Sun, Zimo and Tarnopolsky, Grigory},
  journal={Physical review letters},
  volume={136},
  number={11},
  pages={111602},
  year={2026},
  publisher={APS}
}

@article{Klebanov:2022syt,
    author = "Klebanov, Igor R. and Narovlansky, Vladimir and Sun, Zimo and Tarnopolsky, Grigory",
    title = "{Ginzburg-Landau description and emergent supersymmetry of the (3, 8) minimal model}",
    eprint = "2211.07029",
    archivePrefix = "arXiv",
    primaryClass = "hep-th",
    doi = "10.1007/JHEP02(2023)066",
    journal = "JHEP",
    volume = "02",
    pages = "066",
    year = "2023"
}

@inproceedings{Cordova:2022ruw,
    author = "Cordova, Clay and Dumitrescu, Thomas T. and Intriligator, Kenneth and Shao, Shu-Heng",
    title = "{Snowmass White Paper: Generalized Symmetries in Quantum Field Theory and Beyond}",
    booktitle = "{Snowmass 2021}",
    eprint = "2205.09545",
    archivePrefix = "arXiv",
    primaryClass = "hep-th",
    month = "5",
    year = "2022"
}

@article{Seiberg:2023cdc,
    author = "Seiberg, Nathan and Shao, Shu-Heng",
    title = "{Majorana chain and Ising model -- (non-invertible) translations, anomalies, and emanant symmetries}",
    eprint = "2307.02534",
    archivePrefix = "arXiv",
    primaryClass = "cond-mat.str-el",
    reportNumber = "YITP-SB-2023-14",
    doi = "10.21468/SciPostPhys.16.3.064",
    journal = "SciPost Phys.",
    volume = "16",
    pages = "064",
    year = "2024"
}

@article{Ikhlef2012,
  title = {Integrable Spin Chain for the $\mathrm{SL}(2,R)/\mathbf{U}(1)$ Black Hole Sigma Model},
  author = {Ikhlef, Yacine and Jacobsen, Jesper Lykke and Saleur, Hubert},
  journal = {Phys. Rev. Lett.},
  volume = {108},
  issue = {8},
  pages = {081601},
  numpages = {6},
  year = {2012},
  month = {Feb},
  publisher = {American Physical Society},
  doi = {10.1103/PhysRevLett.108.081601},
  url = {https://link.aps.org/doi/10.1103/PhysRevLett.108.081601}
}

@article{Sinha:2023hum,
    author = "Sinha, Madhav and Yan, Fei and Grans-Samuelsson, Linnea and Roy, Ananda and Saleur, Hubert",
    title = "{Lattice realizations of topological defects in the critical (1+1)-d three-state Potts model}",
    eprint = "2310.19703",
    archivePrefix = "arXiv",
    primaryClass = "hep-th",
    doi = "10.1007/JHEP07(2024)225",
    journal = "JHEP",
    volume = "07",
    pages = "225",
    year = "2024"
}

@article{Douglas:1999vm,
    author = "Douglas, Michael R.",
    title = "{Topics in D geometry}",
    eprint = "hep-th/9910170",
    archivePrefix = "arXiv",
    reportNumber = "RUNHETC-99-40",
    doi = "10.1088/0264-9381/17/5/315",
    journal = "Class. Quant. Grav.",
    volume = "17",
    pages = "1057--1070",
    year = "2000"
}

@article{Polchinski1995,
  title = {Dirichlet Branes and Ramond-Ramond Charges},
  author = {Polchinski, Joseph},
  journal = {Phys. Rev. Lett.},
  volume = {75},
  issue = {26},
  pages = {4724--4727},
  numpages = {0},
  year = {1995},
  month = {Dec},
  publisher = {American Physical Society},
  doi = {10.1103/PhysRevLett.75.4724},
  url = {https://link.aps.org/doi/10.1103/PhysRevLett.75.4724}
}

@misc{Affleck1995conformal,
      title={Conformal Field Theory Approach to the Kondo Effect}, 
      author={Ian Affleck},
      year={1995},
      eprint={cond-mat/9512099},
      archivePrefix={arXiv},
      primaryClass={cond-mat}
}

@article{Kormos:2009sk,
    author = "Kormos, Marton and Runkel, Ingo and Watts, Gerard M. T.",
    title = "{Defect flows in minimal models}",
    eprint = "0907.1497",
    archivePrefix = "arXiv",
    primaryClass = "hep-th",
    reportNumber = "KCL-MTH-09-06",
    doi = "10.1088/1126-6708/2009/11/057",
    journal = "JHEP",
    volume = "11",
    pages = "057",
    year = "2009"
}

@article{Buican:2017rxc,
    author = "Buican, Matthew and Gromov, Andrey",
    title = "{Anyonic Chains, Topological Defects, and Conformal Field Theory}",
    eprint = "1701.02800",
    archivePrefix = "arXiv",
    primaryClass = "hep-th",
    reportNumber = "EFI-16-29, QMUL-17-01",
    doi = "10.1007/s00220-017-2995-6",
    journal = "Commun. Math. Phys.",
    volume = "356",
    number = "3",
    pages = "1017--1056",
    year = "2017"
}

@article{Andrews:1984af,
    author = "Andrews, G. E. and Baxter, R. J. and Forrester, P. J.",
    title = "{Eight vertex SOS model and generalized Rogers-Ramanujan type identities}",
    doi = "10.1007/BF01014383",
    journal = "J. Statist. Phys.",
    volume = "35",
    pages = "193--266",
    year = "1984"
}

@article{Ardonne_2011,
doi = {10.1088/1367-2630/13/4/045006},
url = {https://dx.doi.org/10.1088/1367-2630/13/4/045006},
year = {2011},
month = {apr},
publisher = {},
volume = {13},
number = {4},
pages = {045006},
author = {E Ardonne and J Gukelberger and A W W Ludwig and S Trebst and M Troyer},
title = {Microscopic models of interacting Yang–Lee anyons},
journal = {New Journal of Physics}
}

@article{Feiguin:2006ydp,
    author = "Feiguin, Adrian and Trebst, Simon and Ludwig, Andreas W. W. and Troyer, Matthias and Kitaev, Alexei and Wang, Zhenghan and Freedman, Michael H.",
    title = "{Interacting anyons in topological quantum liquids: The golden chain}",
    eprint = "cond-mat/0612341",
    archivePrefix = "arXiv",
    doi = "10.1103/PhysRevLett.98.160409",
    journal = "Phys. Rev. Lett.",
    volume = "98",
    number = "16",
    pages = "160409",
    year = "2007"
}

@Article{Belletete2023,
author={Bellet{\^e}te, J.
and Gainutdinov, A. M.
and Jacobsen, J. L.
and Saleur, H.
and Tavares, T. S.},
title={Topological Defects in Lattice Models and Affine Temperley--Lieb Algebra},
journal={Communications in Mathematical Physics},
year={2023},
month={Mar},
day={07},
issn={1432-0916},
doi={10.1007/s00220-022-04618-0},
url={https://doi.org/10.1007/s00220-022-04618-0}
}

@book{DiFrancesco:1997nk,
    author = "Di Francesco, P. and Mathieu, P. and Senechal, D.",
    title = "{Conformal Field Theory}",
    doi = "10.1007/978-1-4612-2256-9",
    isbn = "978-0-387-94785-3, 978-1-4612-7475-9",
    publisher = "Springer-Verlag",
    address = "New York",
    series = "Graduate Texts in Contemporary Physics",
    year = "1997"
}

@inproceedings{Saleur:1990uz,
    author = "Saleur, H. and Zuber, J. B.",
    title = "{Integrable lattice models and quantum groups}",
    booktitle = "{Spring School on String Theory and Quantum Gravity (to be followed by 3 day Workshop)}",
    reportNumber = "SACLAY-SPH-T-90-071",
    month = "4",
    year = "1990"
}

@misc{Belletete2018,
      title={Topological defects in lattice models and affine Temperley-Lieb algebra}, 
      author={J. Belletête and A. M. Gainutdinov and J. L. Jacobsen and H. Saleur and T. S. Tavares},
      year={2020},
      eprint={1811.02551},
      archivePrefix={arXiv},
      primaryClass={hep-th}
}

@misc{Belletete2020,
      title={Topological defects in periodic RSOS models and anyonic chains}, 
      author={J. Belletête and A. M. Gainutdinov and J. L. Jacobsen and H. Saleur and T. S. Tavares},
      year={2020},
      eprint={2003.11293},
      archivePrefix={arXiv},
      primaryClass={math-ph}
}

@article{Affleck1995,
    author = "Affleck, Ian",
    editor = "Nowak, Maciej A. and Wegrzyn, Pawel",
    title = "{Conformal field theory approach to the Kondo effect}",
    eprint = "cond-mat/9512099",
    archivePrefix = "arXiv",
    journal = "Acta Phys. Polon. B",
    volume = "26",
    pages = "1869--1932",
    year = "1995"
}

@article{Zamolodchikov1986,
  title={Irreversibility of the Flux of the Renormalization Group in a 2D Field Theory},
  author={Alexander B. Zamolodchikov},
  journal={JETP Letters},
  year={1986},
  volume={43},
  pages={565-567}
}

@article{Petkova:2000ip,
    author = "Petkova, V. B. and Zuber, J. B.",
    title = "{Generalized twisted partition functions}",
    eprint = "hep-th/0011021",
    archivePrefix = "arXiv",
    reportNumber = "UNN-SCM-M-00-07, CERN-TH-2000-322",
    doi = "10.1016/S0370-2693(01)00276-3",
    journal = "Phys. Lett. B",
    volume = "504",
    pages = "157--164",
    year = "2001"
}

@article{Aasen:2020jwb,
  title={Topological defects on the lattice: dualities and degeneracies},
  author={Aasen, David and Fendley, Paul and Mong, Roger SK},
  journal={arXiv preprint arXiv:2008.08598},
  year={2020}
}

@article{Aasen2016,
    author = "Aasen, David and Mong, Roger S. K. and Fendley, Paul",
    title = "{Topological Defects on the Lattice I: The Ising model}",
    eprint = "1601.07185",
    archivePrefix = "arXiv",
    primaryClass = "cond-mat.stat-mech",
    doi = "10.1088/1751-8113/49/35/354001",
    journal = "J. Phys. A",
    volume = "49",
    number = "35",
    pages = "354001",
    year = "2016"
}

@article{Frohlich2004,
  title = {Kramers-Wannier Duality from Conformal Defects},
  author = {Fr\"ohlich, J\"urg and Fuchs, J\"urgen and Runkel, Ingo and Schweigert, Christoph},
  journal = {Phys. Rev. Lett.},
  volume = {93},
  issue = {7},
  pages = {070601},
  numpages = {4},
  year = {2004},
  month = {Aug},
  publisher = {American Phys. Society},
  doi = {10.1103/PhysRevLett.93.070601},
  url = {https://link.aps.org/doi/10.1103/PhysRevLett.93.070601}
}

@article{Grimm2001,
    author = "Grimm, Uwe",
    title = "{Spectrum of a duality twisted Ising quantum chain}",
    eprint = "hep-th/0111157",
    archivePrefix = "arXiv",
    doi = "10.1088/0305-4470/35/3/101",
    journal = "J. Phys. A",
    volume = "35",
    pages = "L25--L30",
    year = "2002"
}

@incollection{Saleur1998,
  title={Lectures on non perturbative field theory and quantum impurity problems},
  author={Saleur, H},
  booktitle={Aspects topologiques de la physique en basse dimension. Topological aspects of low dimensional systems: Session LXIX. 7--31 July 1998},
  pages={473--550},
  year={2002},
  publisher={Springer}
}

@article{Yoshimura2020,
  title = {Non-Hermitian quantum impurity systems in and out of equilibrium: Noninteracting case},
  author = {Yoshimura, Takato and Bidzhiev, Kemal and Saleur, Hubert},
  journal = {Phys. Rev. B},
  volume = {102},
  issue = {12},
  pages = {125124},
  numpages = {23},
  year = {2020},
  month = {Sep},
  publisher = {American Physical Society},
  doi = {10.1103/PhysRevB.102.125124},
  url = {https://link.aps.org/doi/10.1103/PhysRevB.102.125124}
}

@book{Baxter2013,
  title={Exactly Solved Models in Statistical Mechanics},
  author={Baxter, R.J.},
  isbn={9780486318172},
  url={https://books.google.de/books?id=eQzCAgAAQBAJ},
  year={2013},
  publisher={Dover Publications}
}

@article{Fendley1993,
author = {Fendley, P. and Saleur, H. and Zamolodchikov, AL. B.},
title = {Massless flows I: The sine-Gordon and O(n) Models},
journal = {International Journal of Modern Physics A},
volume = {08},
number = {32},
pages = {5717-5750},
year = {1993},
doi = {10.1142/S0217751X93002265},

URL = { 
        https://doi.org/10.1142/S0217751X93002265
    
},
eprint = { 
        https://doi.org/10.1142/S0217751X93002265
    
}
}

@article{tartaglia2016fused,
  title={Fused RSOS lattice models as higher-level nonunitary minimal cosets},
  author={Tartaglia, Elena and Pearce, Paul A},
  journal={Journal of Physics A: Mathematical and Theoretical},
  volume={49},
  number={18},
  pages={184002},
  year={2016},
  publisher={IOP Publishing}
}

@article{pearce2025unitary,
  title={Unitary and Nonunitary ADE minimal models: Coset graph fusion algebras, defects, entropies, SREEs and dilogarithm identities},
  author={Pearce, Paul A and Heymann, Jared and Quella, Thomas},
  journal={arXiv preprint arXiv:2512.21808},
  year={2025}
}

@Article{Belletete:2018eua,
author={Bellet{\^e}te, J.
and Gainutdinov, A. M.
and Jacobsen, J. L.
and Saleur, H.
and Tavares, T. S.},
title={Topological Defects in Lattice Models and Affine Temperley-Lieb Algebra},
journal={Communications in Mathematical Physics},
year={2023},
day={01},
volume={400},
number={2},
pages={1203-1254},
issn={1432-0916},
    eprint = "1811.02551",
    archivePrefix = "arXiv",
    primaryClass = "hep-th",
doi={10.1007/s00220-022-04618-0},
url={https://doi.org/10.1007/s00220-022-04618-0}
}

@article{barkeshli2019symmetry,
  title={Symmetry fractionalization, defects, and gauging of topological phases},
  author={Barkeshli, Maissam and Bonderson, Parsa and Cheng, Meng and Wang, Zhenghan},
  journal={Physical Review B},
  volume={100},
  number={11},
  pages={115147},
  year={2019},
  publisher={APS}
}

@article{buican2022galois,
  title={Galois orbits of TQFTs: symmetries and unitarity},
  author={Buican, Matthew and Radhakrishnan, Rajath},
  journal={Journal of High Energy Physics},
  volume={2022},
  number={1},
  pages={4},
  year={2022},
  publisher={Springer}
}

@article{etingof2005fusion,
  title={On fusion categories},
  author={Etingof, Pavel and Nikshych, Dmitri and Ostrik, Viktor},
  journal={Annals of mathematics},
  pages={581--642},
  year={2005},
  publisher={JSTOR}
}

@article{Lootens:2019xjv,
    author = "Lootens, Laurens and Vanhove, Robijn and Haegeman, Jutho and Verstraete, Frank",
    title = "{Galois Conjugated Tensor Fusion Categories and Nonunitary Conformal Field Theory}",
    eprint = "1902.11241",
    archivePrefix = "arXiv",
    primaryClass = "quant-ph",
    doi = "10.1103/PhysRevLett.124.120601",
    journal = "Phys. Rev. Lett.",
    volume = "124",
    number = "12",
    pages = "120601",
    year = "2020"
}

@article{brody2014biorthogonal,
  title={Biorthogonal quantum mechanics},
  author={Brody, Dorje C},
  journal={Journal of Physics A: Mathematical and Theoretical},
  volume={47},
  number={3},
  pages={035305},
  year={2014},
  publisher={IOP Publishing}
}

@article{Gils_2009,
	doi = {10.1103/physrevlett.103.070401},
  
	url = {https://doi.org/10.1103%2Fphysrevlett.103.070401},
  
	year = 2009,
	month = {aug},
  
	publisher = {American Physical Society ({APS})},
  
	volume = {103},
  
	number = {7},
  
	author = {Charlotte Gils and Eddy Ardonne and Simon Trebst and Andreas W. W. Ludwig and Matthias Troyer and Zhenghan Wang},
  
	title = {Collective States of Interacting Anyons, Edge States, and the Nucleation of Topological Liquids},
  
	journal = {Physical Review Letters}
}

@article{Koo:1993wz,
    author = "Koo, W. M. and Saleur, H.",
    title = "{Representations of the Virasoro algebra from lattice models}",
    eprint = "hep-th/9312156",
    archivePrefix = "arXiv",
    reportNumber = "USC-93-025, YCTP-P22-93",
    doi = "10.1016/0550-3213(94)90018-3",
    journal = "Nucl. Phys. B",
    volume = "426",
    pages = "459--504",
    year = "1994"
}

@inbook{Kirillov:1991ec,
  author    = {A. N. Kirillov and N. Yu. Reshetikhin},
  title     = {Representations of the Algebra $U_q(sl(2))$, $q$-Orthogonal Polynomials and Invariants of Links},
  booktitle = {New Developments in the Theory of Knots},
  pages     = {202--256},
  year      = {1991},
  doi       = {10.1142/9789812798329_0012},
  url       = {https://www.worldscientific.com/doi/abs/10.1142/9789812798329_0012}
}

@article{Itzykson_1986,
doi = {10.1209/0295-5075/2/2/004},
url = {https://doi.org/10.1209/0295-5075/2/2/004},
year = {1986},
month = {jul},
publisher = {},
volume = {2},
number = {2},
pages = {91},
author = {C. Itzykson and H. Saleur and J.-B. Zuber},
title = {Conformal Invariance of Nonunitary 2d-Models},
journal = {Europhysics Letters}
}
\bibliographystyle{apsrev4-2}

\end{document}